\documentclass{article}

\usepackage{booktabs}
\usepackage{amsfonts}
\usepackage{nicefrac}
\usepackage{microtype}
\usepackage[dvipsnames]{xcolor}
\usepackage{amsmath,amsthm,amsfonts}
\usepackage{algorithm}
\usepackage{algorithmic}
\usepackage{amssymb}
\usepackage{multirow}
\usepackage{xspace}
\usepackage{arydshln}
\usepackage{graphicx}
\usepackage{wrapfig}
\usepackage{placeins}
\usepackage{subcaption}
\usepackage{graphicx}
\usepackage{soul}
\usepackage{pifont}
\usepackage{tikz}

\newcommand{\ours}{SPICE\xspace}
\newcommand{\oursbold}{\textbf{\ours}\xspace}
\newcommand{\gasloc}{GASLoC\xspace}
\newcommand{\choco}{CHOCO-SGD\xspace}

\newcommand{\diloco}{DiLoCo\xspace}
\newcommand{\dadam}{DAdam\xspace}
\newcommand{\allreduce}{All-Reduce\xspace}

\newcommand{\sparseloco}{SparseLoCo\xspace}
\newcommand{\topk}{\textsc{Top}-\(k\)\xspace}

\definecolor{lavanda}{HTML}{E89A72}
\definecolor{arancia}{RGB}{215,154,2}
\definecolor{islandA}{HTML}{E89A72}
\definecolor{islandB}{RGB}{146,112,171}
\definecolor{islandC}{RGB}{70,70,70}
\definecolor{slowedLink}{HTML}{A03B33}
\definecolor{rosso}{HTML}{914735}
\definecolor{verde}{HTML}{96ad91}

\newcommand{\eg}{e.g.,\xspace}
\newcommand{\cmark}{{\color{verde}\ding{51}}}
\newcommand{\xmark}{{\color{lavanda}\ding{55}}}

\usepackage{microtype}
\usepackage{graphicx}
\usepackage{booktabs} %

\usepackage{hyperref}
\usepackage{cleveref}
\crefname{figure}{Figure}{Figures}
\Crefname{figure}{Figure}{Figures}
\crefname{table}{Table}{Tables}
\Crefname{table}{Table}{Tables}

\usepackage[preprint]{mlsys2025}

\mlsystitlerunning{\ours: Sparse Parameter Gossip for  LLM Pretraining}

\begin{document}

\twocolumn[
\mlsystitle{\ours: Sparse Parameter Gossip for  LLM Pretraining}

\begin{mlsysauthorlist}
\mlsysauthor{Pietro Cagnasso}{conc,mila}
\mlsysauthor{Daniele Kopyshevskiy}{sorb}
\mlsysauthor{Edouard Oyallon}{sorb,cnrs}
\mlsysauthor{Eugene Belilovsky}{conc,mila}
\end{mlsysauthorlist}

\mlsysaffiliation{conc}{Concordia University}
\mlsysaffiliation{mila}{Mila}
\mlsysaffiliation{sorb}{Sorbonne Université}
\mlsysaffiliation{cnrs}{CNRS}

\mlsyscorrespondingauthor{}{}

\mlsyskeywords{Machine Learning, MLSys}

\vskip 0.3in

\begin{abstract}
Pretraining language models across geographically distributed resources requires communication-efficient methods that operate under limited and variable internet bandwidth.
Decentralized training replaces global collectives with peer exchanges, but full-model messages make larger peer neighborhoods costly.
We introduce \oursbold, which combines sparse parameter gossip with error feedback through corrected parameter values that remain directly interpretable as peer neighborhoods change.
These sparse exchanges make broader neighborhoods affordable, achieving better training quality with less communication than the evaluated dense decentralized baselines. As the replica count grows, \ours increases the number of peers contacted while keeping individual messages sparse, limiting the quality degradation observed with dense gossip.
Under bandwidth constraints, \ours achieves higher training throughput than DDP and dense gossip, while sustaining throughput during transient link slowdowns.

\end{abstract}
]

\printAffiliationsAndNotice{}  %

\section{Introduction}
\label{sec:introduction}

Training large language models at scale relies on distributing computation across many model replicas, each potentially spanning several GPUs. In standard synchronous data parallel training, these replicas average their gradients before each optimizer step~\cite{li2020pytorchdistributed}, typically through an \allreduce collective~\cite{patarasuk2009bandwidth}. Each step therefore requires communication among all replicas and waits for the slowest participant, making training efficiency dependent on the connectivity between them.

Even within tightly connected GPU clusters, communication bandwidth can limit training efficiency, motivating low-precision transfers to reduce communication volume~\cite{deepseekai2025deepseekv3}.
These constraints become more severe when training extends over geographically distributed data centers, where bandwidth can be limited and vary across participants.
Recent large language model pretraining runs over the internet~\cite{jaghouar2024intellect1,lidin2026covenant72b} demonstrate the opportunity to pool resources across sites, but exploiting these resources efficiently requires limiting the cost of communication~\citep{nabli2024acco,kopyshvskiy2026lacco}.

\begin{figure*}[!t]
    \centering
    \begingroup
\definecolor{commGray}{HTML}{484C50}
\definecolor{commNode}{HTML}{b3b3b3}
\definecolor{commRingNode}{HTML}{2F2926}
\definecolor{commRing}{HTML}{D0D1D4}
\definecolor{commAllReduceEdge}{HTML}{B8BABF}
\definecolor{commPurple}{HTML}{8F6CA3}
\definecolor{commPurpleLight}{HTML}{B196C0}
\definecolor{commPurpleOther}{HTML}{EADFEF}
\definecolor{commRed}{HTML}{E89A72}
\definecolor{commRedLight}{HTML}{E89A72}
\definecolor{commRedOther}{HTML}{F8E6BC}
\definecolor{commTrack}{HTML}{ECEDEF}

\def\commEdgesA{0/13,13/2,2/8,8/14,14/4,4/12,12/6,6/10,
  10/1,1/9,9/15,15/5,5/11,11/3,3/7,7/0}
\def\commEdgesB{0/3,3/9,9/4,4/15,15/7,7/12,12/2,2/10,
  10/14,14/6,6/1,1/5,5/13,13/8,8/11,11/0}
\def\commNextEdgesA{0/5,5/9,9/3,3/13,13/7,7/12,12/6,6/10,
  10/1,1/8,8/15,15/4,4/14,14/2,2/11,11/0}
\def\commNextEdgesB{0/6,6/1,1/5,5/11,11/4,4/9,9/15,15/3,
  3/12,12/8,8/13,13/2,2/7,7/14,14/10,10/0}

\begin{tikzpicture}[x=1cm,y=1cm,
  every node/.style={inner sep=0pt,outer sep=0pt},
  line cap=round,line join=round]
  \path[use as bounding box] (-2.05,-1.85) rectangle (15.25,2.12);

  \foreach \x in {2.2,6.6,11.0}
    \draw[commRing,line width=.45pt] (\x,-1.80)--(\x,2.07);

  \foreach \panel/\title/\subtitle/\volume in {
    0/{Ring All-Reduce}/{global collective}/{\nicefrac{2(R-1)}{R}\,d=1.875d},
    1/{GASLoC}/{$q=2\cdot\mathrm{dense}$}/{2d},
    2/{\textbf{\ours}}/{$q=2\cdot\delta=6.25\%$}/{q\delta d=0.125d},
    3/{\textbf{\ours}}/{$q=4\cdot\delta=3.125\%$}/{q\delta d=0.125d}}
  {
    \begin{scope}[xshift=\panel*4.4cm]
      \ifcase\panel
        \def\commColor{commRingNode}\def\commFill{commAllReduceEdge}
        \def\commFraction{.9375}
      \or
        \def\commColor{commPurple}\def\commFill{commPurpleLight}
        \def\commOther{commPurpleOther}
        \def\commFraction{1}
      \or
        \def\commColor{commRed}\def\commFill{commRedLight}
        \def\commOther{commRedOther}
        \def\commFraction{.0625}
      \or
        \def\commColor{commRed}\def\commFill{commRedLight}
        \def\commOther{commRedOther}
        \def\commFraction{.0625}
      \fi

      \node[font=\normalfont\normalsize,text=black] at (0,1.91) {\title};
      \node[font=\small,text=commGray] at (0,1.55) {\subtitle};

      \ifnum\panel=0
        \def\commSnapshots{0/0/{\text{for every }t}}
        \def\commRadius{.98}
      \else
        \def\commSnapshots{0/-1.02/t,1/1.02/{t+1}}
        \def\commRadius{.90}
      \fi
      \foreach \step/\offset/\round in \commSnapshots {
      \begin{scope}[xshift=\offset cm]
      \node[font=\footnotesize,text=commGray] at (0,-1.18) {$\round$};
      \ifnum\step=0
        \let\commActiveA\commEdgesA
        \let\commActiveB\commEdgesB
        \def\commNeighborsA{13,7}\def\commNeighborsB{3,11}
      \else
        \let\commActiveA\commNextEdgesA
        \let\commActiveB\commNextEdgesB
        \def\commNeighborsA{5,11}\def\commNeighborsB{6,10}
      \fi
      \foreach \i in {0,...,15} {
        \coordinate (v\i) at ({90-22.5*\i}:\commRadius);
      }

      \ifnum\panel=0
        \foreach \i in {0,...,15}{
          \pgfmathtruncatemacro{\j}{mod(\i+1,16)}
          \draw[commAllReduceEdge,line width=3pt] (v\i)--(v\j);
        }
      \else
        \foreach \i in {0,...,15}{
          \pgfmathtruncatemacro{\j}{mod(\i+1,16)}
          \draw[commRing,line width=.35pt] (v\i)--(v\j);
        }
        \ifnum\panel=1
          \def\commWidth{2.8pt}
        \else
          \ifnum\panel=2
            \def\commWidth{1pt}
          \else
            \def\commWidth{.5pt}
          \fi
        \fi
        \foreach \a/\b in \commActiveA
          \draw[\commOther,line width=\commWidth]
            (v\a)--(v\b);
        \ifnum\panel=3
          \foreach \a/\b in \commActiveB
            \draw[\commOther,line width=\commWidth]
              (v\a)--(v\b);
        \fi
        \foreach \j in \commNeighborsA
          \draw[\commFill,line width=\commWidth] (v0)--(v\j);
        \ifnum\panel=3
          \foreach \j in \commNeighborsB
            \draw[\commFill,line width=\commWidth] (v0)--(v\j);
        \fi
      \fi

      \foreach \i in {0,...,15} {
        \ifnum\panel=0
          \fill[commRingNode] (v\i) circle[radius=.036];
        \else
          \fill[commNode] (v\i) circle[radius=.036];
        \fi
      }
      \ifnum\panel>0
        \foreach \i in \commNeighborsA
          \fill[\commColor] (v\i) circle[radius=.055];
        \ifnum\panel=3
          \foreach \i in \commNeighborsB
            \fill[\commColor] (v\i) circle[radius=.055];
        \fi
      \fi
      \node[rectangle,fill=\commColor,minimum size=.15cm] at (v0) {};
      \end{scope}
      }

      \fill[commTrack] (-1.96,-1.49) rectangle (1.96,-1.39);
      \fill[\commFill] (-1.96,-1.49)
        rectangle ({-1.96+3.92*\commFraction},-1.39);
      \node[anchor=west,text=commGray,font=\small]
        at (-1.96,-1.73) {$V=\volume$};
    \end{scope}
  }
\end{tikzpicture}
\endgroup
    \caption{
        \textbf{Communication patterns and outbound communication volume per worker for 16 replicas.} \gasloc and \ours support changing neighborhoods across communication rounds, illustrated at $t$ and $t+1$, whereas Ring \allreduce implements a global collective over a fixed ring. The square marks a reference worker, whose peer connections are emphasized. For a model with $d$ parameters, Ring \allreduce transmits $1.875d$ values per worker, while dense two-peer \gasloc transmits $2d$ values. Instead, \ours can send selected parameter values to two peers at $6.25\%$ density or four peers at $3.125\%$ density, both totaling $q\delta d=0.125d$. %
        \ours combines dynamic neighborhoods with sparse messages, allowing a fixed value budget to be allocated between message density and peer count.
    }
    \label{fig:communication-patterns}
\end{figure*}
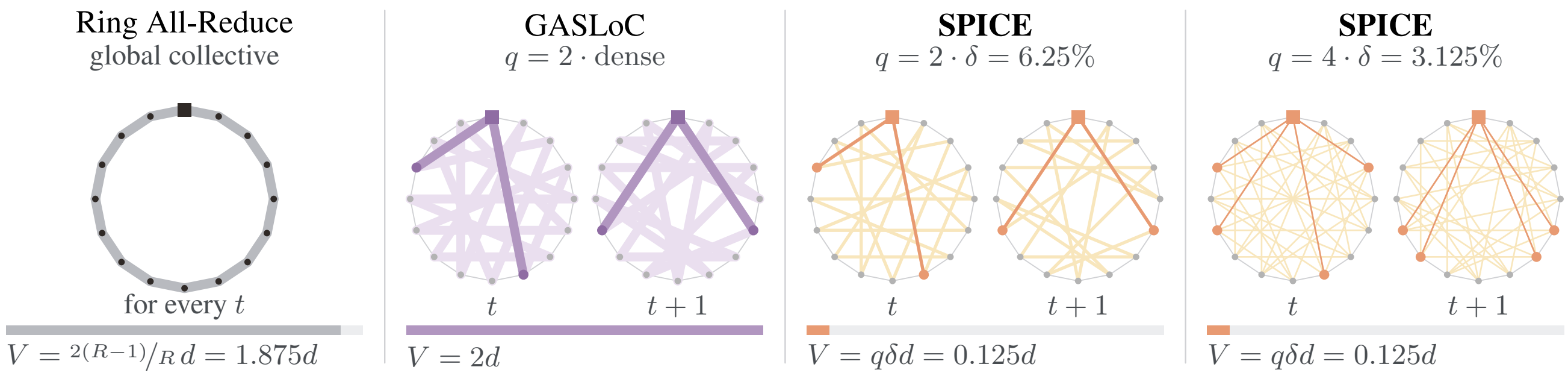

Decentralized (gossip) training replaces the typical global \allreduce operation with exchanges among selected peers~\cite{koloskova2020unified}. This has several potential advantages as it can avoid communication bottlenecks, support heterogeneous bandwidth, and maintain robustness to failure of individual links. %
However, combining gossip methods with adaptive optimizers has historically been difficult, limiting their applicability to demanding settings such as pretraining foundational models.
Recent works by \citet{wang2024promise} and \citet{cagnasso2026unifying} have shown that decentralized methods can indeed achieve strong performance, %
narrowing the training quality gap with synchronous baselines in LLM pretraining.
In decentralized training, communication topology also affects convergence through its spectral properties~\cite{nabli2023acid}, while \gasloc empirically shows that communicating with more peers can improve training quality~\cite{cagnasso2026unifying}.
With full model exchanges, however, each additional peer increases communication volume, motivating smaller messages that allow broader peer coverage at a lower communication cost (\Cref{fig:communication-patterns}).

One approach to reducing communication is compression, through sparsification or quantization. Sparsification transmits only a subset of gradient or pseudogradient coordinates, selected, for example, through chunked \topk selection~\cite{sarfi2026sparseloco}. The resulting information loss can be mitigated through error feedback, which retains omitted update components for later communication~\cite{stich2021error}. Applying these mechanisms to parameter gossip is challenging because replicas evolve from different models: a compressed update relative to the sender's model cannot generally be applied directly to the receiver's model. Methods such as \choco address this mismatch by reconstructing communicated states from accumulated compressed updates on a fixed communication graph~\cite{koloskova2019decentralized}, but changing peers requires making the relevant reconstruction state available to new receivers.
After a local update, sparsification lets a sender transmit selected parameter values directly, reducing message size while allowing any receiver to interpret them without a shared reference model.
This motivates sparse parameter gossip that combines directly interpretable messages with the accumulated correction from error feedback.

The framework of \citet{cagnasso2026unifying} provides a natural way to realize this design: it expresses gossip as a local pseudo-gradient formed from differences between a node's parameters and those received from its peers. This formulation allows us to apply a sparse update with error feedback locally, then communicate the resulting parameter values at the selected coordinates. We use this mechanism to introduce \oursbold, a decentralized training algorithm that combines chunked \topk sparsification with error feedback while retaining the Nesterov outer optimizer for consensus acceleration. The transmitted parameter values incorporate accumulated error corrections and can be interpreted directly by any receiving peer, without persistent reconstruction caches.

\begin{figure}[tb]
    \centering
    \includegraphics[width=\linewidth]{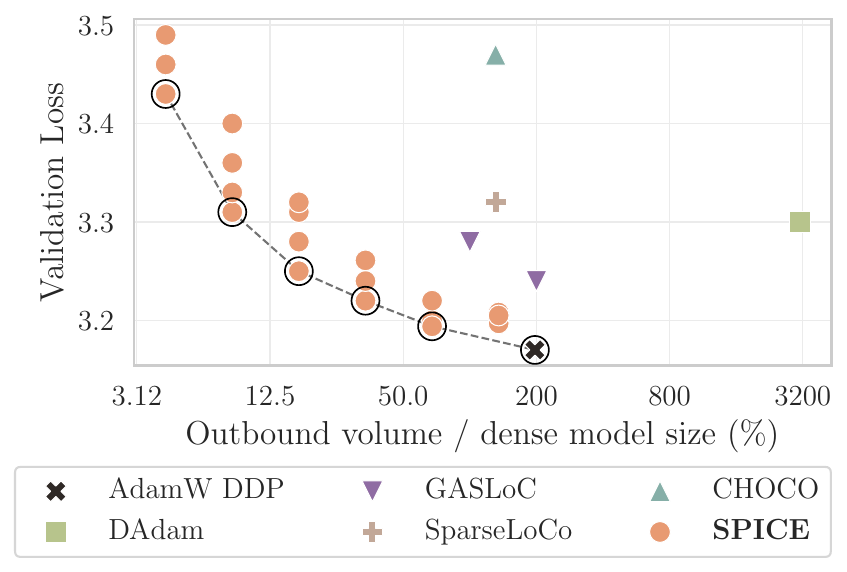}
    \caption{\textbf{Quality–communication tradeoff for decentralized LLM pretraining.} Final validation loss for a 134M-parameter model trained on 2.70B tokens with 32 replicas. \ours configurations vary peer count and sparsity. Outbound volume includes transmitted values and 12-bit packed indices and is normalized by the size of a dense model. Circled points and the dashed line indicate the empirical Pareto frontier among evaluated configurations. \ours, achieving lower validation loss than all evaluated decentralized baselines, lies on the empirical Pareto frontier.}
    \label{fig:pareto-comm-vol}
\end{figure}

Our contributions are:
\begin{enumerate}
    \item We propose \ours, a decentralized training algorithm that integrates sparsification and error memory into parameter gossip through corrected absolute model endpoints.
    \item We empirically show that \ours achieves \emph{lower validation loss} than evaluated decentralized baselines, while transmitting less data (\Cref{fig:pareto-comm-vol}).
    \item We show that \ours \emph{reduces end-to-end training time} to a fixed token budget across several bandwidth caps compared with dense \gasloc and DDP, and \emph{retains most of the training throughput} under transient bandwidth reductions, including severe slowdowns.
\end{enumerate}

\section{Related Works}

\paragraph{Decentralized language-model training.}
Decentralized training replaces global aggregation with exchanges among selected peers, making the communication graph and peer selection rule central to information propagation~\cite{boyd2006randomized,nabli2025decentralized,nabli2023dadao}.
D-PSGD combines neighbor averaging with stochastic optimization~\cite{lian2017can}, while Stochastic Gradient Push and subsequent analyses extend decentralized training to directed or time-varying graphs and multiple local updates between exchanges~\cite{assran2019stochastic,koloskova2020unified}.
For transformer training, \dadam studies local adaptive optimization with parameter mixing~\cite{wang2024promise}.
\gasloc builds on this approach for language-model pretraining by combining local adaptive updates, peer neighborhoods defined by a shared random permutation at each round, and an outer optimizer with momentum to accelerate consensus~\cite{cagnasso2026unifying}.
These parameter exchanges allow replicas with different models to aggregate their peers' endpoints, but still transmit a full model to each peer.
\ours adopts the same neighborhood construction as \gasloc while reducing message size through sparse exchanges of corrected model endpoints.

\paragraph{Local optimization and compressed updates.}
Federated averaging and Local SGD reduce the overall communication volume by performing multiple local optimization steps between aggregation rounds~\cite{mcmahan2017communication,stich2019localsgd}.
FedOpt generalizes model averaging by applying an outer optimizer to aggregated model differences used as pseudo-gradients~\cite{reddi2021adaptive}.
\diloco applies this inner-outer optimization framework for language-model pretraining with local AdamW updates and an outer Nesterov optimizer~\cite{douillard2023diloco}.
\gasloc connects this framework to parameter gossip and recovers \diloco under exact global averaging~\cite{cagnasso2026unifying}.
Compression complements local updates by reducing the size of each message rather than the frequency of communication.
Error feedback compensates for omitted information by retaining it for subsequent communication rounds~\cite{stich2018sparsified,karimireddy2019error}.
Beyond compensating for compression error, error feedback can be combined with explicit momentum: EF21-SGDM establishes improved convergence guarantees by adding Polyak momentum to EF21~\cite{fatkhullin2023momentum}, while DeMo uses error feedback to compress communicated momentum in language-model training~\cite{peng2024demo}.
QSparse-local-SGD combines local updates with sparsification, quantization, and error compensation, but uses SGD without an outer momentum optimizer~\cite{basu2019qsparse}.
\sparseloco studies this combination in language-model pretraining with local AdamW updates, using chunked \topk compression of pseudo-gradients and error feedback as a surrogate for the outer momentum~\cite{sarfi2026sparseloco}.
In SparseLoCo, replicas compute their pseudo-gradients from a shared round-start model, so averaging these displacements is equivalent to averaging their corresponding endpoints.
In parameter gossip, replicas generally start from different models, so applying a peer’s displacement to the receiver’s model does not recover the intended peer’s endpoint.
\ours addresses this difference by transmitting selected absolute endpoint values that incorporate the accumulated error correction.

\paragraph{Compressed gossip.}
WASH communicates selected weights among models through random peer-to-peer permutations, keeping the models compatible with final weight averaging~\cite{fournier2024wash}.
Compressed gossip methods instead combine neighboring models during training, with \choco reconstructing model estimates by accumulating compressed differences~\cite{koloskova2019decentralized,koloskova2020decentralized}. Each message therefore depends on the receiver’s reconstruction history, so the standard formulation assumes a fixed communication graph to ensure that receivers retain the history needed.
\ours instead averages selected absolute endpoint values that incorporate error memory and require no receiver reconstruction history.

\paragraph{Communication topology.}
Beyond message compression, topology design controls the communication cost of propagating information among replicas.
MATCHA samples graph matchings to balance communication cost and convergence~\cite{wang2019matcha}, while one-peer exponential graphs and EquiTopo provide sparse communication schemes with favorable mixing properties~\cite{ying2021exponential,song2022communication}.
However, faster consensus alone does not establish better training quality, as the spectral gap can fail to predict decentralized learning performance under identically distributed data~\cite{vogels2023beyond}.

\section{Method} \label{sec:method}

\subsection{\ours}

\paragraph{Sparse Parameter Gossip Challenges.}
\label{sec:method:parameter-gossip}
Combining gossip with gradient compression and error feedback is challenging because these methods operate on model differences. It thus requires specifying the reference model of each transmitted update.
Consider $R$ replicas with $d$ parameters and identical initializations.
At round $t$, replica $i$ starts from $x_i^{(t)}\in\mathbb{R}^d$, performs a local optimizer step to obtain $y_i^{(t)}$, and computes its local displacement $\Delta_i^{(t)}=x_i^{(t)}-y_i^{(t)}$.
For example, a local-step-then-mix form of decentralized SGD is
\begin{equation}
\begin{aligned}
    y_i^{(t)} &= x_i^{(t)}-\eta_t g_i^{(t)},\\
    x_i^{(t+1)} &= \sum_{j=1}^{R} w_{ij}^{(t)}y_j^{(t)},
\end{aligned}
\label{eq:dense-gossip-motivation}
\end{equation}
where $g_i^{(t)}$ is a local stochastic gradient, $\eta_t$ is the learning rate, and the nonnegative mixing weights sum to one and are nonzero only for the replica itself and its current peers~\cite{koloskova2020unified}.
This exchange averages model parameters.
Replicas generally start a round from different models, so applying a sender's displacement $\Delta_j^{(t)}$ to the receiver's model $x_i^{(t)}$ does not recover $y_j^{(t)}$.
Compressing these displacements alone therefore does not directly reproduce parameter gossip.

CHOCO-SGD addresses this mismatch by maintaining a shared estimate $\hat{x}_j^{(t)}$ of each sender's model and compressing the difference from that estimate~\cite{koloskova2019decentralized,koloskova2020decentralized}.
Using the local-step-then-gossip ordering, its updates can be written as
\begin{equation}
\begin{aligned}
    m_i^{(t)} &= \mathcal{C}\!\left(y_i^{(t)}-\hat{x}_i^{(t)}\right),\\
    \hat{x}_i^{(t+1)} &= \hat{x}_i^{(t)}+m_i^{(t)},\\
    x_i^{(t+1)} &= y_i^{(t)}
      +\gamma\sum_{j=1}^{R}w_{ij}
        \bigl(\hat{x}_j^{(t+1)}-\hat{x}_i^{(t+1)}\bigr),
\end{aligned}
\label{eq:choco-motivation}
\end{equation}
where $\mathcal{C}$ is a compression operator, $\gamma$ is the consensus stepsize, and $W=(w_{ij})$ is a fixed symmetric doubly stochastic mixing matrix.
The sender and its neighbors apply the same transmitted increment $m_i^{(t)}$ to maintain matching estimates.
Information omitted by compression remains in the difference between the model and its estimate and can enter subsequent messages.

Its memory-efficient formulation avoids storing each neighbor estimate by maintaining their weighted sum $s_i^{(t)}=\sum_j w_{ij}\hat{x}_j^{(t)}$, requiring only three model-sized states.
This formulation makes the dependence on a fixed graph explicit: the accumulated state $s_i$ encodes a specific weighted neighborhood, and changing that neighborhood generally requires information that cannot be recovered from the stored sum alone.

\paragraph{Sparse Parameters as Local Update Directions.}
Our proposed method \ours, instead, transmits selected absolute parameter values, combining the endpoint exchange of \gasloc with sparsification and error feedback~\cite{cagnasso2026unifying,sarfi2026sparseloco}.
To see the basic mechanism, first omit error feedback.
When replica $j$ transmits coordinate $y_j^{(t)}[c]$, receiver $i$ uses its own round-start parameter as the reference and forms
\begin{equation}
\begin{aligned}
    &x_i^{(t)}[c]-y_j^{(t)}[c]\\
    &\quad=\underbrace{x_i^{(t)}[c]-x_j^{(t)}[c]}_{\text{model disagreement}}
      +\underbrace{\Delta_j^{(t)}[c]}_{\text{local displacement}}.
\end{aligned}
\label{eq:receiver-reference}
\end{equation}
This direction is relative to the receiver's model and includes both the sender's local displacement and the disagreement term needed for parameter gossip.
As in \gasloc, averaging these differences over the available contributions gives a pseudo-gradient for an outer optimizer that provides consensus acceleration \cite{cagnasso2026unifying}.
Absolute parameter values require neither a shared starting model nor a stored estimate of the sender's previous model, allowing neighborhoods to change between rounds without transferring model estimates to new peers.
As detailed below, this ability is central to combining a small communication budget with mixing across the replicas.

To complete this construction, we must specify which coordinates are sent, how their values incorporate retained error, and how to average messages with different supports.

\paragraph{Error-Feedback and Sparsification.}
\label{sec:method:sparsification}
We select coordinates using the local displacements and their retained history.
Each replica maintains a residual buffer $e_i^{(t)}$, initialized with $e_i^{(0)}=0$.
At round $t$, it combines this buffer with the current local pseudo-gradient,
\begin{equation}
    \tilde{e}_i^{(t)}=\beta e_i^{(t)}+\Delta_i^{(t)},
    \label{eq:memory-accumulator}
\end{equation}
where $\beta\in[0,1]$ controls the decay of retained history.

We partition the flattened accumulator into contiguous chunks of $b$ coordinates and select the $k$ entries with largest absolute values within each chunk.
This chunked \topk rule follows ~\cite{sarfi2026sparseloco,peng2024demo} and assigns the same selection quota to each full chunk of a tensor, with density $\delta=k/b$.
Let $\mathcal{M}_i^{(t)}\in\{0,1\}^d$ denote the resulting mask and $\mathcal{I}_i^{(t)}=\{c:\mathcal{M}_i^{(t)}[c]=1\}$ the set of selected coordinates.
The residual retained for the next round is
\begin{equation}
\begin{aligned}
    e_i^{(t+1)} &= (\mathbf{1}-\mathcal{M}_i^{(t)})\odot \tilde{e}_i^{(t)},
\end{aligned}
    \label{eq:memory-selection}
\end{equation}
where $\odot$ denotes elementwise multiplication.

Unselected components remain in the buffer and enter subsequent accumulators, allowing repeated displacements to accumulate rather than being discarded after each round.
The mask determines which coordinates are transmitted, while $\mathcal{M}_i^{(t)}\odot\tilde e_i^{(t)}$ is the selected accumulated displacement used to construct the corrected endpoints below.

\paragraph{Correcting Endpoints with Error Feedback.}
\label{sec:method:corrected-endpoints}
On selected coordinates, the accumulator contains the current local displacement and the retained historical correction.
Sending only $y_i^{(t)}[\mathcal{I}_i^{(t)}]$ would omit this correction.
We therefore construct a corrected endpoint
\begin{equation}
    z_i^{(t)}=y_i^{(t)}
    -\mathcal{M}_i^{(t)}\odot\beta e_i^{(t)}.
    \label{eq:corrected-endpoint}
\end{equation}
For a selected coordinate $c\in\mathcal{I}_i^{(t)}$, this is equivalent to
\begin{equation}
    z_i^{(t)}[c]=x_i^{(t)}[c]-\tilde e_i^{(t)}[c].
    \label{eq:selected-endpoint}
\end{equation}
Replica $i$ sends $(\mathcal{I}_i^{(t)},z_i^{(t)}[\mathcal{I}_i^{(t)}])$ to its peers.
On unselected coordinates, $z_i^{(t)}[c]=y_i^{(t)}[c]$, so the self endpoint retains the current local update.

The accumulated displacement is encoded as $x_i^{(t)}[c]-z_i^{(t)}[c]$ and cleared from the residual on selected coordinates, while remaining in the buffer on unselected coordinates.
Together, these two cases give
\begin{equation}
\begin{aligned}
    \tilde{e}_i^{(t)}&=\underbrace{\beta e_i^{(t)}+\Delta_i^{(t)}}_{\text{accumulated displacement}}\\
    &=\underbrace{e_i^{(t+1)}}_{\text{retained in memory}}+\underbrace{\mathcal{M}_i^{(t)}\odot\left(x_i^{(t)}-z_i^{(t)}\right)}_{\text{encoded in transmitted endpoints}}.
\end{aligned}
    \label{eq:local-error-conservation}
\end{equation}
Thus, any historical correction cleared from memory is incorporated into the transmitted endpoint values.
We next combine the full self endpoint with the selected peer values through coordinate-wise averaging.

\paragraph{Randomized Coordinate-Wise Gossip and Outer Optimization.}
\label{sec:method:gossip-outer}
In methods such as \choco, messages depend on accumulated reconstruction state, which must be available when introducing a new peer. The corrected absolute endpoints in \ours can instead be interpreted by any receiver without previous exchanges. This allows peer neighborhoods to change between rounds. We use randomized peer selection, motivated by analyses showing that randomized one- and two-peer exchanges can mix efficiently as the number of replicas grows~\cite{cagnasso2026unifying}. At each round, a newly sampled shared permutation determines the neighborhoods, and each replica sends its selected endpoint values to $q$ peers.

Each message contains a fraction $\delta$ of the model coordinates, giving $q\delta d$ transmitted parameter values per round (\cref{fig:communication-patterns}). \choco can match this payload using a fixed graph with the same peer count and density. However, sparse fixed topologies such as rings can limit mixing and hinder training as the number of replicas grows~\cite{nabli2023acid}. The advantage of \ours is that it combines sparse messages with changing communication partners without transferring reconstruction state. Relative to the complete-graph \choco baseline used in our experiments, contacting only $q$ peers reduces outbound volume by a factor of $(R-1)/q$ at equal message density.

Since peers can select different coordinates, the contributors to coordinate $c$ are
\begin{equation}
    \mathcal{P}_{i,c}^{(t)}=\{i\}\cup
    \{j\in\mathcal{N}_i^{(t)}:c\in\mathcal{I}_j^{(t)}\}.
    \label{eq:coordinate-contributors}
\end{equation}
We average their endpoint values and form the outer residual,
\begin{equation}
\begin{aligned}
    G_i^{(t)}[c] =x_i^{(t)}[c]- \frac{1}{|\mathcal{P}_{i,c}^{(t)}|}
      \sum_{j\in\mathcal{P}_{i,c}^{(t)}}z_j^{(t)}[c].
\end{aligned}
    \label{eq:coordinate-gossip}
\end{equation}
The self endpoint always contributes, including coordinates outside $\mathcal{I}_i^{(t)}$.
Peers that do not transmit a coordinate contribute neither a value nor a count to its average.
If no peer transmits $c$, then $G_i^{(t)}[c] = x_i^{(t)}[c] - z_i^{(t)}[c]$, and the outer residual retains the local displacement and any selected historical correction.

To separate the effects of local optimization and model disagreement, one can define the displacement $D_j^{(t)}=x_j^{(t)}-z_j^{(t)}=\Delta_j^{(t)}+\mathcal{M}_j^{(t)}\odot\beta e_j^{(t)}$.
The outer residual can then be written as
\begin{equation}
\begin{aligned}
    G_i^{(t)}[c]
    &=\underbrace{x_i^{(t)}[c]
      -\frac{\sum_{j\in\mathcal{P}_{i,c}^{(t)}}x_j^{(t)}[c]}
            {|\mathcal{P}_{i,c}^{(t)}|}}_{\text{consensus disagreement}}\\
    &\quad+\underbrace{\frac{\sum_{j\in\mathcal{P}_{i,c}^{(t)}}D_j^{(t)}[c]}
                 {|\mathcal{P}_{i,c}^{(t)}|}}_{\text{local displacements}}.
\end{aligned}
    \label{eq:gossip-decomposition}
\end{equation}
The first term corrects disagreement between the receiver's starting model and the neighbors' starting models.
The second averages their local displacements and selected historical corrections.
When the contributing replicas share a starting model, the disagreement term vanishes and gossip reduces to averaging these corrected displacements.
Otherwise, coordinate selection affects both displacement aggregation and consensus.

Following \gasloc~\cite{cagnasso2026unifying}, we apply an outer Nesterov momentum optimizer to $G_i^{(t)}$, starting from $x_i^{(t)}$.
This choice follows the connection between outer momentum and accelerated consensus developed for dense gossip by \citet{cagnasso2026unifying}.
Algorithm~\ref{alg:sparse-gasloc} summarizes this training procedure.

\begin{algorithm}[t]
\caption{\ours}
\label{alg:sparse-gasloc}
{\footnotesize
\begin{algorithmic}[1]
\STATE \textbf{Input:} initialize $x_1^{(0)} = \dots = x_R^{(0)}$, initialize $e_i^{(0)} = 0~\forall i$,
$\mathrm{Optimizer}^{\mathrm{in}}$, $\mathrm{Optimizer}^{\mathrm{out}}$, EF memory decay $\beta$, and $q$, $k$, $b$

\FOR{$t=0,1,\dots,T-1$}
    \STATE Sample a shared permutation $\sigma^{(t)}$ of $\{1,\dots,R\}$
    \FOR{Worker $i\in\{1,\dots,R\}$ in parallel}
        \item[] {$\triangleright$ \textbf{Inner update}}
        \STATE $y_i^{(t)} \gets \mathrm{Optimizer}^{\mathrm{in}}\!
            \left(x_i^{(t)},~\nabla f(x_i^{(t)})\right)$
        \STATE $\Delta_i^{(t)} \gets x_i^{(t)} - y_i^{(t)}$
\item[] {\color{lavanda}{$\triangleright$ \textbf{Error-feedback support selection}}}
        \STATE {\color{lavanda}$\tilde e_i^{(t)} \gets \beta e_i^{(t)} + \Delta_i^{(t)}$}
        \STATE {\color{lavanda}$\mathcal{M}_i^{(t)}, \mathcal{I}_i^{(t)}
            \gets \mathrm{Top-}k\left(\tilde e_i^{(t)}\right)$}
        \STATE {\color{lavanda}$z_i^{(t)} \gets y_i^{(t)} - \mathcal{M}_i^{(t)} \odot \beta e_i^{(t)}$}
        \STATE {\color{lavanda}$e_i^{(t+1)} \gets (\mathbf{1} - \mathcal{M}_i^{(t)}) \odot \tilde{e}_i^{(t)}$}
\item[] {$\triangleright$ \textbf{Gossip step}}
        \STATE $\mathcal{N}_i^{(t)} \gets$
            \textsc{sample\_peers}{$\left(\sigma^{(t)},~i\right)$}
        \STATE exchange $\left(\mathcal{I}_i^{(t)},~
            z_i^{(t)}[\mathcal{I}_i^{(t)}]\right)$
            with all $j \in \mathcal{N}_i^{(t)}$
        \FOR{$c \in \{1,\dots,d\}$}
            \STATE $\mathcal{P}_{i,c}^{(t)}
                \gets \{i\} \cup
                \{j \in \mathcal{N}_i^{(t)} :
                    c \in \mathcal{I}_j^{(t)}\}$
            \STATE $G_i^{(t)}[c]
                \gets x_i^{(t)}[c]
                - \frac{1}{|\mathcal{P}_{i,c}^{(t)}|}
                  \sum_{j \in \mathcal{P}_{i,c}^{(t)}}
                      z_j^{(t)}[c]$
        \ENDFOR
        \item[] {$\triangleright$ \textbf{Outer update}}
        \STATE $x_i^{(t+1)} \gets \mathrm{Optimizer}^{\mathrm{out}}\!
            \left(x_i^{(t)},~G_i^{(t)}\right)$
    \ENDFOR
\ENDFOR
\STATE \textbf{Output:}
    $\bar{x}^{(T)}=\frac{1}{R}\sum_{i=1}^R x_i^{(T)}$
\end{algorithmic}
}
\end{algorithm}

With full support (\topk$=b$) and zero initial residual, the residual remains zero and $z_i^{(t)}=y_i^{(t)}$ at every round, recovering dense \gasloc.%

\subsection{Communication Cost}
\label{sec:method:communication-cost}

\label{sec:method:implementation}
Each worker saves its round-start parameters to compute the displacement used for support selection and the pseudo-gradient used by the outer optimizer.
For selection, the trainable local parameter shards are viewed as one vector and partitioned into chunks of $b$ coordinates.
Chunks can cross PyTorch's parameter boundaries, so participating workers have to use a consistent parameter ordering and shard layout.

The selected coordinates become a message containing two buffers: their indices and their parameter values.
For $b\le4096$, an index needs only 12 bits to identify a position within its chunk; thus, two indices can be packed into three bytes.
Since each chunk contributes the \topk entries in order, the receiver infers the chunk from the entry's position in the message.

Given this message format, let $d_{\mathrm{loc}}$ be the number of trainable parameters in a local shard and $q$ the number of peers.
Assuming complete chunks and packed index pairs, each worker selects $s=(k/b)d_{\mathrm{loc}}=\delta d_{\mathrm{loc}}$ coordinates and sends the same message to its $q$ peers.
If each value occupies $v$ bytes and the indices add 1.5 bytes per selected coordinate, then
\begin{equation}
    V_{\mathrm{send}}
    =q\delta d_{\mathrm{loc}}\left(v+1.5\right).
    \label{eq:sparse-payload-bytes}
\end{equation}
Dense exchange with the same peers sends $qvd_{\mathrm{loc}}$ bytes, so the payload ratio is $\delta(1+3/(2v))$.

\begin{table}[t]
    \centering
    \caption{\textbf{Comparison of communication and optimization capabilities.} Methods comparison: \cmark\xspace and \xmark\xspace indicate supported and unsupported features, respectively. \ours brings together peer-to-peer communication, communication acceleration, sparse communication, and dynamic neighborhoods in a unified framework.}
    \label{tab:methods-cmp}
    \resizebox{\linewidth}{!}{{\fontsize{8}{8}\selectfont
    \begin{tabular}{lcccc}
        \toprule
        & \multirow{2}{*}{\textbf{P2P}} & \multirow{2}{*}{\textbf{Acceleration}} & \multirow{2}{*}{\textbf{Sparse}} & \textbf{Dynamic}\\
        \textbf{Method} &&&& \textbf{Neighborhoods} \\
        \midrule
         AdamW DDP & \xmark & \xmark & \xmark & \xmark \\
         \dadam & \cmark & \xmark & \xmark & \cmark \\
         \gasloc & \cmark & \cmark & \xmark & \cmark \\
         \sparseloco & \xmark & \xmark & \cmark & \xmark \\
         \choco & \cmark & \xmark & \cmark & \xmark \\
         \ours & \cmark & \cmark & \cmark & \cmark\\
         \bottomrule
    \end{tabular}
    }}
\end{table}

\section{Experimental Results}\label{sec:exps}

\begin{table*}[!t]
    \centering
    \caption{\textbf{Final validation loss.} (a) Validation loss for the 134M model trained on 2.70B tokens with \(R\) replicas. (b) Validation loss for the 551M model trained on 11.01B tokens across \(R\) replicas. DDP and \sparseloco communicate through global collectives (gc), while the other methods use peer-to-peer communication. \dadam and \choco use a complete communication graph, while \gasloc and \ours sample a time-varying graph. Methods with sparse communication use density \(\delta=3.125\%\), with \(k=128\) and block size \(b=4096\). We highlight in bold the best \textbf{decentralized} result for each replica count and underline the best result among \underline{sparse} methods. In case of a tie, the method with the lowest communication volume is highlighted. \ours’s advantage over the evaluated decentralized baselines grows with the replica count, as its expanding neighborhood limits loss degradation.}
    \label{tab:main-results}
    \begin{subtable}[t]{0.55\textwidth}
    \centering
    \caption{134M parameters, 2.70B tokens.}
    \label{tab:main-results-134m}
    \fontsize{7}{8.5}\selectfont
    \setlength{\tabcolsep}{2pt}
    \begin{tabular}{lcccccc}
         \toprule
         \textbf{Method} & \textbf{Peers} $q$ & $\mathbf{R=8}$ & $\mathbf{R=16}$ & $\mathbf{R=32}$ & $\mathbf{R=64}$ & $\mathbf{R=128}$ \\
         \midrule
         AdamW DDP (\textit{ref.}) & gc & 3.17 & 3.17 & 3.17 & 3.17 & 3.17 \\
         \dadam & $R-1$ & 3.23 & 3.25 & 3.30 & 3.36 & 3.42 \\
         \gasloc & 1 & 3.20 & 3.23 & 3.28 & 3.35 & 3.44 \\
         \gasloc & 2 & 3.19 & 3.21 & 3.24 & 3.31 & 3.37 \\
         \sparseloco & gc & 3.27 & 3.28 & 3.32 & 3.38 & 3.42 \\
         \choco & $R-1$ & 3.27 & 3.35 & 3.47 & 3.67 & 3.86 \\
         \oursbold \textbf{(ours)} & $\nicefrac{R}{2}$ & \underline{\textbf{3.19}} & \underline{\textbf{3.19}} & \underline{\textbf{3.20}} & \underline{\textbf{3.21}} & \underline{\textbf{3.23}} \\
         \bottomrule
    \end{tabular}
    \end{subtable}\hfill
    \begin{subtable}[t]{0.43\textwidth}
    \centering
    \caption{551M parameters, 11.01B tokens.}
    \label{tab:main-results-551m}
    \fontsize{7}{8.5}\selectfont
    \setlength{\tabcolsep}{2pt}
    \begin{tabular}{lcccc}
         \toprule
         \textbf{Method} & \textbf{Peers} $q$ & $\mathbf{R=8}$ & $\mathbf{R=16}$ & $\mathbf{R=32}$ \\
         \midrule
         AdamW DDP (\textit{ref.}) & gc & 2.63 & 2.63 & 2.63 \\
         \dadam & $R-1$ & 2.70 & 2.73 & 2.77 \\
         \gasloc & 1 & 2.67 & 2.71 & 2.75 \\
         \gasloc & 2 & \textbf{2.64} & 2.68 & 2.73 \\
         \sparseloco & gc & 2.71 & 2.74 & 2.78 \\
         \choco & $R-1$ & 2.73 & 2.81 & 2.95 \\
         \oursbold \textbf{(ours)} & $\nicefrac{R}{2}$ & \underline{2.65} & \underline{\textbf{2.68}} & \underline{\textbf{2.70}} \\
         \bottomrule
    \end{tabular}
    \end{subtable}
\end{table*}

\begin{table}[t]
    \centering
    \caption{\textbf{7B Mixture of Experts.} Final validation loss and downstream length-normalized accuracy for the 7B MoE trained on 26B tokens across 8 replicas. \ours uses \topk = 256 and block size \(b=4096\) for a density $\delta=6.25\%$. Compared with dense two-peer \gasloc, \ours achieves comparable model quality with \(8\times\) less communication volume.}
    \label{tab:moe}
    \resizebox{\linewidth}{!}{
        \begin{tabular}{lcccc}
            \toprule
            \textbf{Method} & \textbf{Comm. Vol.} & \textbf{Loss} & \textbf{HSwag} & \textbf{PIQA} \\
            \midrule
            AdamW DDP (\textit{ref.}) & $1.75d$ & 2.362 & \(52.71 \pm 0.49\) & \(72.36 \pm 1.04\) \\
            \dadam-complete & $7d$ & 2.421 & \(48.87 \pm 0.49\) & \(71.71 \pm 1.05\) \\
            \gasloc ($q=1$) & $d$ & 2.370 & \(51.93 \pm 0.50\) & \(71.60 \pm 1.04\) \\
            \gasloc ($q=2$) & $2d$ & 2.367 & \(52.56 \pm 0.50\) & \(72.58 \pm 1.05\) \\
            \oursbold \textbf{(ours)} ($q=4$) & $0.25d$ & 2.368 & \(52.38 \pm 0.50\) & \(72.42 \pm 1.05\) \\
            \bottomrule
        \end{tabular}
    }
\end{table}

\subsection{Experimental Setup}
We train two dense Llama-3-style decoder-only Transformers~\cite{grattafiori2024llama} with 134M and 551M parameters, and the mixture-of-experts model OLMoE-1B-7B~\cite{muennighoff2025olmoe}, on FineWeb~\cite{penedo2024fineweb}.
Their training budgets are 2.70B, 11.01B, and 26.05B tokens, respectively, corresponding to approximately 20 tokens per parameter for the dense models~\cite{hoffmann2022training} and per active parameter for OLMoE.
The data are tokenized with the LLaMA-2 tokenizer~\cite{touvron2023llama2} and packed into sequences of 2048 tokens, with a global batch of 1024 sequences, or approximately 2M tokens, shared across replicas.
All methods perform local updates with AdamW, with the peak learning rate tuned separately for each method and a schedule consisting of warmup followed by cosine decay.
Its second-order momentum $\beta_2$ is scaled according to \citet{belilovsky2026how}.
Where applicable, we also tune the outer learning rate and Nesterov momentum.

\paragraph{Baselines.} We compare \ours with AdamW DDP, \sparseloco, \dadam, \gasloc, and \choco, with replicas communicating after every local optimizer update in all decentralized methods.
AdamW DDP provides the reference for training with exact gradient aggregation at every update~\cite{li2020pytorchdistributed}.
We choose \dadam as an adaptive decentralized baseline evaluated on transformer training~\cite{wang2024promise}.
We also compare with \gasloc using one or two peers per round, which provides the dense counterpart of \ours and measures the quality and communication cost achievable through peer selection alone~\cite{cagnasso2026unifying}.
Finally, \choco provides a baseline for compressed decentralized communication~\cite{koloskova2020decentralized}.
We evaluate it on the complete graph because it relies on a static communication graph, a dependence explicit in its memory-efficient formulation. Fixed sparse graphs could match the peer count of our randomized selections (\eg a ring for $q=2$), but would impose a substantial mixing disadvantage, potentially slowing convergence~\cite{nabli2023acid}.
We evaluate SparseLoCo at the same sparsity density as the other sparse baselines, without quantization or outer momentum, and with \(H=1\).
\Cref{tab:methods-cmp} summarizes the key distinctions among these methods.

\paragraph{Implementation.}
We evaluate training quality with a simulator of local updates and message exchanges, allowing us to study up to 128 replicas without requiring 128 physical GPUs.
We measure execution timings separately using a TorchTitan implementation~\cite{liang2025torchtitan}.
This implementation packs the selected indices within each chunk into 12 bits each, storing two indices in three bytes, and exchanges the packed indices and selected parameter values through actual peer-to-peer communication.
Values retain the parameter \texttt{dtype} and are exchanged in batches after the packed indices are sent.
The receiver then unpacks the indices and accumulates the received values at the corresponding coordinates.

\begin{figure*}[t]
    \centering
    \begin{subfigure}[t]{0.49\textwidth}
        \centering
        \includegraphics[width=\linewidth]{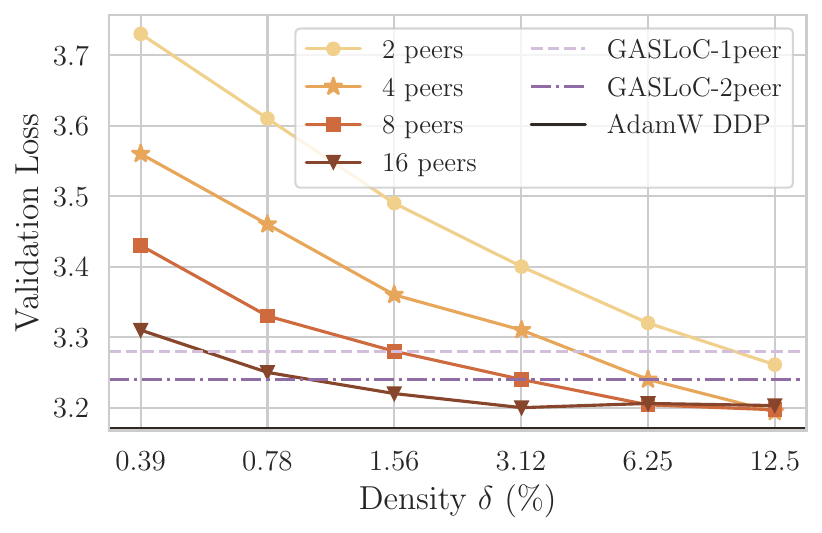}
        \caption{Joint effect of density and peer count.}
        \label{fig:density}
    \end{subfigure}\hfill
    \begin{subfigure}[t]{0.49\textwidth}
        \centering
        \includegraphics[width=\linewidth]{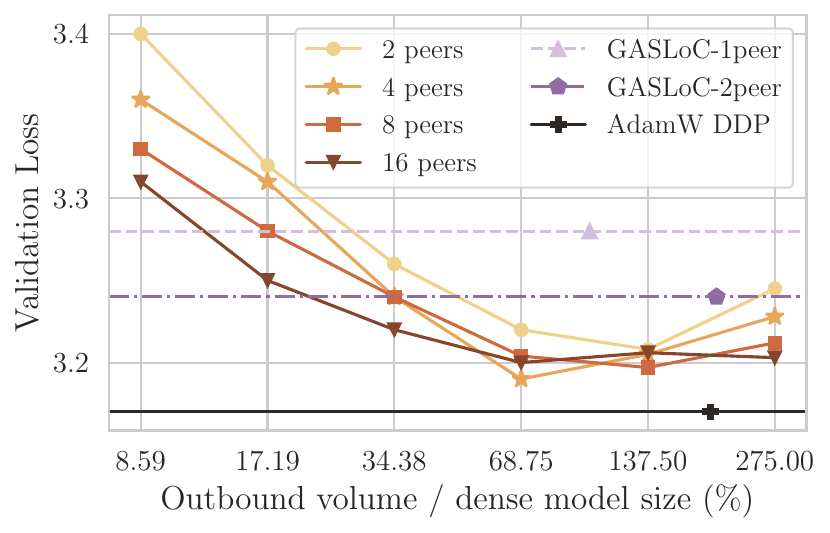}
        \caption{Allocation of a fixed communication budget.}
        \label{fig:volume}
    \end{subfigure}
    \caption{\textbf{Effect of density and peer count on validation loss.} The 134M model is trained with 32 replicas for 2.70B tokens.
    (a) shows the joint effect of density and peer count, the two factors that determine the total communication volume. %
    (b) compares validation loss at fixed outbound communication volume, %
    showing the trade-off between more peers and more coordinates per peer under a fixed budget.
    Increasing density brings diminishing improvements when more peers are used, while configurations with fewer peers remain more sensitive to sparsification.
    At small communication budgets, %
    sending sparse messages to more peers yields lower validation losses.
    }
    \label{fig:density-volume}
\end{figure*}

\subsection{Training Results}
For the 134M model, \ours achieves the best validation loss among both the sparse and decentralized methods evaluated in \cref{tab:main-results-134m}.
Increasing the replica count from 8 to 128 raises validation loss by only 1.3\% for \ours, compared with 5.6\% for dense two-peer \gasloc.
This reduces a known difficulty of methods using inner and outer optimizers, whose performance can degrade as the replica count increases~\cite{horoi2026model,belilovsky2026how}.
Each message contains only 3.125\% of the parameter coordinates, transmitting more than 30 times fewer parameter values than a dense decentralized exchange.
This sparsity allows us to expand the neighborhood proportionally to the replica count, using $q=R/2$, at a fraction of the volume required for dense exchanges with the same peers.
As the fixed global batch is divided among more replicas, the neighborhood grows by the same factor that the local batch size decreases, keeping the total number of tokens processed by the communicating peers constant.
For each coordinate, however, only peers that select it contribute to the outer gradient, so this does not imply constant token coverage at every coordinate.
The token coverage maintained across the neighborhood provides a possible explanation for the reduced degradation, which results in a validation loss advantage of 0.14 over dense two-peer \gasloc at 128 replicas.

Using the same density and neighborhood rule, \ours retains the best validation loss among the evaluated sparse methods when the model size increases from 134M to 551M parameters, as shown in \cref{tab:main-results-551m}.
As with the 134M model, its performance is comparable to dense two-peer \gasloc at smaller replica counts, with an increasing advantage as the replica count grows.
From 8 to 32 replicas, its validation loss increases by 1.9\%, compared with 3.4\% for the dense variant, extending the reduced degradation to a larger model and training budget.

Table~\ref{tab:moe} extends the evaluation to OLMoE-1B-7B~\cite{muennighoff2025olmoe}, a mixture-of-experts model with 7B parameters. With 8 replicas, \ours exchanges 6.25\% of the parameter coordinates with four peers per round, using one-eighth the communication volume than dense two-peer \gasloc. It achieves a nearly identical validation loss and comparable accuracy in HellaSwag~\cite{zellers2019hellaswag} and PIQA~\cite{bisk2019piqa}, with both accuracy differences falling within the reported error bars. These results show that the communication savings also extend to different and larger model architectures while preserving competitive training quality.

\subsection{Effect of Density and Peer Count}
\label{sec:results:density}

Peer count $q$ and density $\delta$ are the two main controls of communication volume in \ours.
Their product $q\delta$ determines the total number of parameter values sent per round relative to one dense model exchange, with the corresponding byte cost given in Section~\ref{sec:method:communication-cost}.
The same communication volume can therefore be allocated to more peers or more coordinates per peer.
Figure~\ref{fig:density} compares how these choices affect validation loss.

With fewer peers, increasing density reduces validation loss throughout the tested range.
With more peers, loss instead levels off once density reaches a few percent, and the advantage of larger neighborhoods decreases as density increases.
Larger neighborhoods are therefore most beneficial under aggressive sparsification.
This is consistent with sparse messages from more peers partly compensating for the limited number of coordinates in each message.
The resulting exchanges may already provide enough information that transmitting additional coordinates brings little improvement, even while communication volume remains below that of a dense model exchange.

Figure~\ref{fig:volume} compares these configurations at equal outbound communication volume, including the packed indices.
At smaller communication budgets, allocating the same volume to more peers reduces validation loss.
With sixteen peers, \ours improves on both dense \gasloc baselines while sending less than half the bytes required for one dense model exchange per round.

\subsection{Training Under Limited Bandwidth}
\label{sec:results:timings}

Training over the internet often relies on connections with substantially less bandwidth than datacenter interconnects~\cite{wang2023cocktailsgd,jaghouar2024intellect1,lidin2026covenant72b}.
In this setting, communication can limit training speed, while bandwidth stragglers and temporary link failures can further delay synchronization.
These constraints motivate measuring how reducing communication volume affects training time.

\subsubsection{End-to-End Training Time}

\begin{figure*}[t]
    \centering
    \begin{subfigure}[t]{0.49\textwidth}
        \centering
        \includegraphics[width=\linewidth]{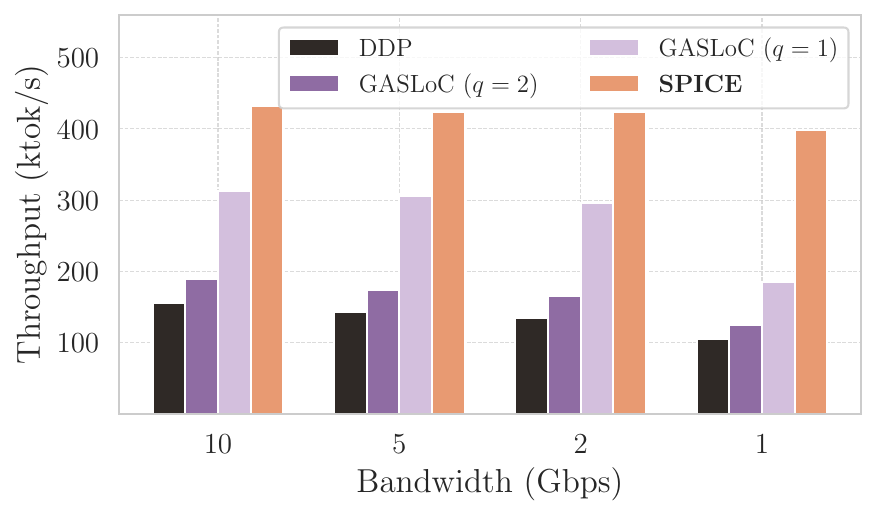}
        \caption{134M parameters.}
        \label{fig:training-speed-134m}
    \end{subfigure}\hfill
    \begin{subfigure}[t]{0.49\textwidth}
        \centering
        \includegraphics[width=\linewidth]{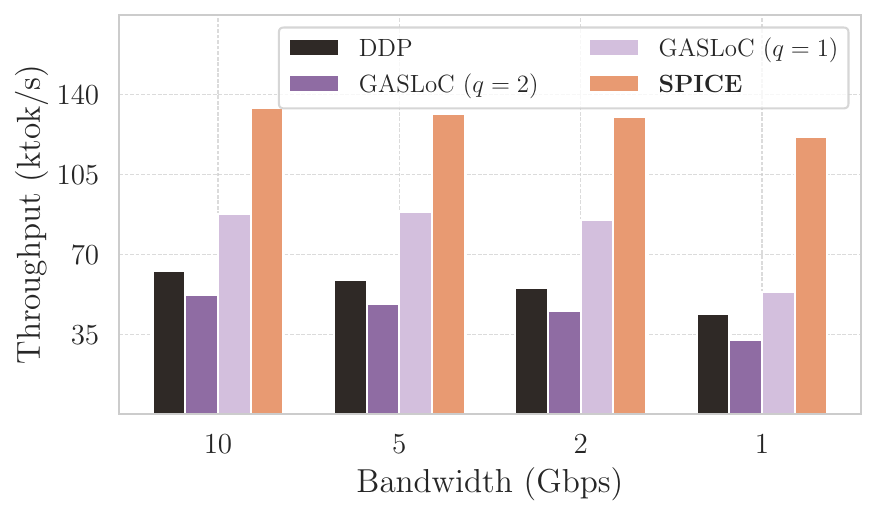}
        \caption{551M parameters.}
        \label{fig:training-speed-551m}
    \end{subfigure}
    \caption{\textbf{Training throughput under limited bandwidth.} Aggregate throughput in thousands of tokens per second for the 134M and 551M models on eight RTX A6000 GPUs, with the same bandwidth cap applied separately to each replica's links. Higher is better.
    \ours uses four peers, \topk = 128, and block size $b=4096$, corresponding to density $\delta=3.125\%$.
    \ours achieves the highest throughput at every tested bandwidth and retains nearly constant throughput as the cap decreases from 10 to 1\,Gbps, while DDP and both dense \gasloc variants lose throughput over this range.}
    \label{fig:training-speed}
\end{figure*}

We benchmark the 134M and 551M models on eight NVIDIA RTX A6000 GPUs within a single machine, with one training replica per GPU, using our TorchTitan implementation~\cite{liang2025torchtitan} with actual peer-to-peer communication.
For \ours, each exchange transmits the selected parameter values and their 12-bit packed indices. 
The timings therefore include index packing, transmission of both indices and values, and unpacking and accumulation at the receiver.
To control bandwidth, we place each replica in a separate Linux network namespace and connect the namespaces through a virtual Ethernet bridge.
We use \texttt{tc netem} to cap outgoing traffic at the replica interface and incoming traffic at the corresponding interface on the host.
The specified rate therefore limits each replica's aggregate sending and receiving bandwidth separately, and is shared by all its peer exchanges.
We apply the same rate to all eight replicas and evaluate caps of 1, 2, 5, and 10\,Gbps.
NCCL uses its socket transport over these interfaces, with direct GPU transfers, shared memory transport, and InfiniBand disabled to prevent communication from bypassing the caps.
These runs impose bandwidth limits without injecting additional latency or link failures.
The uncapped setting instead uses NCCL without the namespace bandwidth controls.

Figure~\ref{fig:training-speed} shows that \ours achieves the highest training throughput for both model sizes at every tested bandwidth.
Its throughput changes little as bandwidth decreases from 10 to 1\,Gbps, while all dense methods slow down over this range, increasing the advantage of sparsification under tighter bandwidth limits.
This advantage holds even though \ours communicates with four peers, compared with one or two for dense \gasloc.
With \topk = 128 and $b=4096$, each sparse message contains only $3.125\%$ of the model coordinates, allowing \ours to reach more peers while sending less data overall, including the coordinate indices.
The resulting throughput shows that sparsification can support a larger peer count without incurring the communication cost of dense exchanges.

\begin{figure}[!t]
    \centering
    \includegraphics[width=\linewidth]{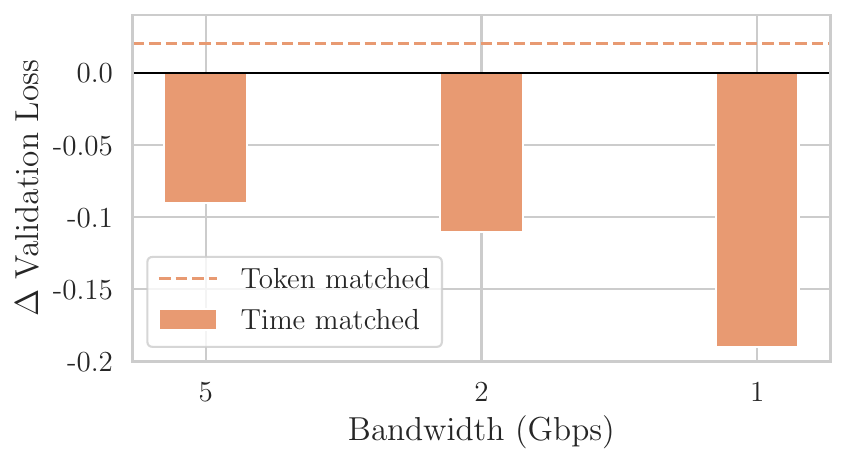}
    \caption{\textbf{Time-matched training under limited bandwidth.} Validation loss differences relative to DDP for \ours (\(q=4\), \(\delta=3.125\%\)), training the 134M model on 8 A6000 GPUs with uniform bandwidth caps set using \texttt{tc}.
    At equal training time, \ours outperforms DDP at all bandwidth caps shown with a reduction of 6\% in the slowest setting.}
    \label{fig:time-matched}
\end{figure}

To assess how this throughput advantage affects training quality, we use the time required by DDP to complete its training budget as the reference.
At each bandwidth cap, we train \ours for more steps to match this elapsed time, allowing it to process more tokens.
\Cref{fig:time-matched} shows that \ours achieves lower validation loss on the 134M model under this comparison, with a reduction of 6\% at 1\,Gbps.
The additional steps enabled by sparse communication therefore improve validation loss within a fixed time window.

\begin{figure*}[t]
    \centering
    \settoheight{\dimen0}{\includegraphics[width=0.35\textwidth]{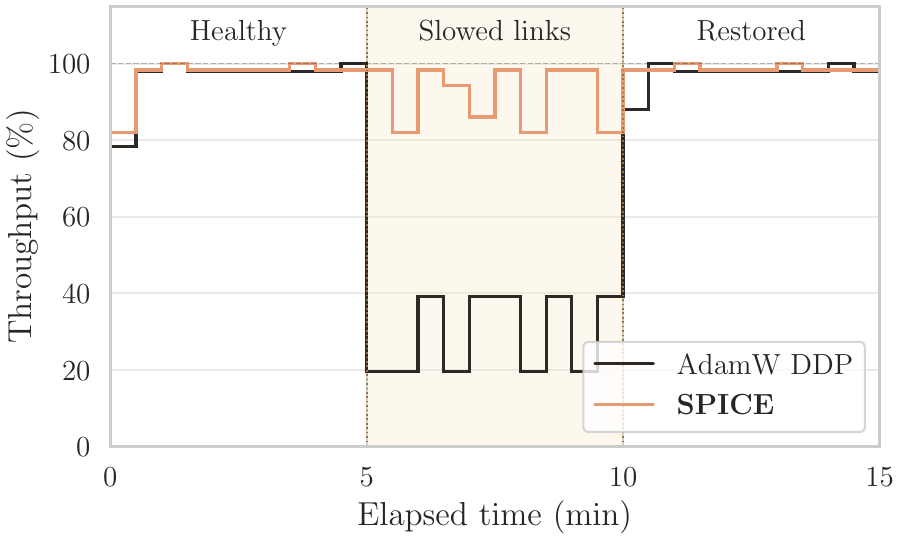}}
    \edef\slowLinksPanelHeight{\the\dimen0}
    \begin{subfigure}[t]{0.28\textwidth}
        \centering
        \vspace{0pt}
        \resizebox*{!}{\slowLinksPanelHeight}{%
\begin{tikzpicture}[
    font=\small,
    worker/.style={
        circle, draw=black!75, fill=white,
        minimum size=6mm, inner sep=0pt
    },
    healthy/.style={
        black!25, line width=0.5pt
    },
    slowed/.style={
        slowedLink, densely dashed, line width=0.8pt
    }
]
    \draw[draw=islandA, fill=islandA!10, thick,
          rotate around={22.5:(-0.68,1.8)}]
        (-0.68,1.8) ellipse[x radius=2.30cm,y radius=1.05cm];

    \draw[draw=islandB, fill=islandB!10, thick,
          rotate around={67.5:(1.8,-0.68)}]
        (1.8,-0.68) ellipse[x radius=2.30cm,y radius=1.05cm];

    \draw[draw=islandC, fill=islandC!10, thick,
          rotate around={135:(-1.63,-1.63)}]
        (-1.63,-1.63) ellipse[x radius=1.50cm,y radius=0.75cm];

    \foreach \i in {1,...,8} {
        \pgfmathsetmacro{\angle}{157.5-45*(\i-1)}
        \coordinate (p\i) at (\angle:2.5cm);
    }

    \foreach \i in {1,...,7} {
        \pgfmathtruncatemacro{\next}{\i+1}
        \foreach \j in {\next,...,8} {
            \ifnum\i<4
                \ifnum\j<4
                    \draw[healthy] (p\i)--(p\j);
                \else
                    \ifnum\j>6
                        \draw[healthy] (p\i)--(p\j);
                    \fi
                \fi
            \else
                \draw[healthy] (p\i)--(p\j);
            \fi
        }
    }

    \foreach \i in {1,2,3} {
        \foreach \j in {4,5,6} {
            \draw[slowed] (p\i)--(p\j);
        }
    }

    \foreach \i in {1,...,8} {
        \node[worker] at (p\i) {\i};
    }

    \node[text=islandA, font=\large]
        at (-1.4,3.2) {Island A};
    \node[text=islandB, anchor=west, font=\large]
        at (2.55,-1.7) {Island B};
    \node[text=islandC, anchor=east, font=\large]
        at (-2.1,-2.65) {Island C};

    \draw[slowed] (-1.1,-3.3)--(-0.4,-3.3);
    \node[anchor=west, font=\large] at (-0.3,-3.3) {Slowed links};
\end{tikzpicture}%
        }
        \caption{Island topology.}
        \label{fig:islands}
    \end{subfigure}\hfill
    \begin{subfigure}[t]{0.35\textwidth}
        \centering
        \vspace{0pt}
        \includegraphics[height=\slowLinksPanelHeight]{figures/throughput_500mbps.pdf}
        \caption{500\,Mbps.}
        \label{fig:slow-links-500mbps}
    \end{subfigure}\hfill
    \begin{subfigure}[t]{0.35\textwidth}
        \centering
        \vspace{0pt}
        \includegraphics[height=\slowLinksPanelHeight]{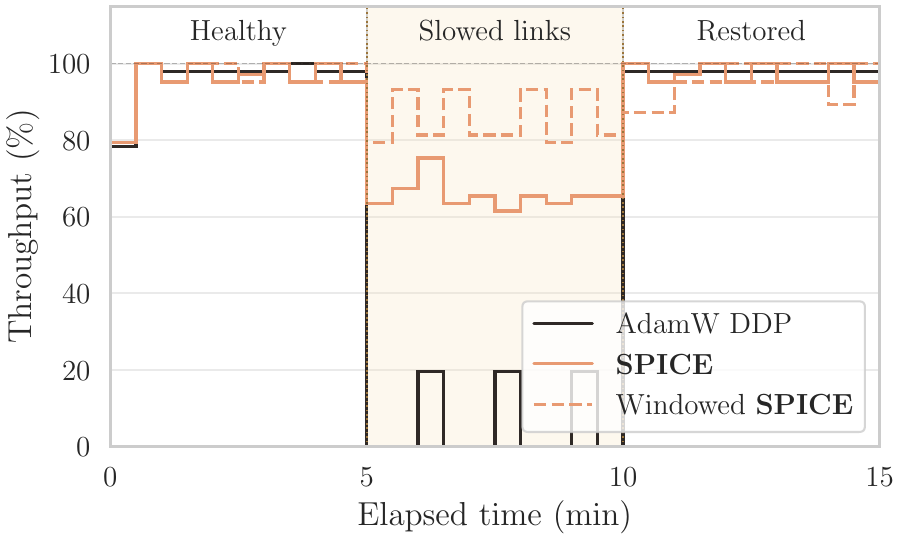}
        \caption{100\,Mbps.}
        \label{fig:slow-links-100mbps}
    \end{subfigure}
    \caption{\textbf{Training throughput with slow links between communication islands.} Eight replicas of the 134M model run on eight NVIDIA RTX A6000 GPUs, with one replica per GPU.
    (a) Replicas are grouped into islands A, B, and C containing three, three, and two replicas, respectively. Dashed edges mark the affected links between A and B.
    (b, c) Throughput when each directed worker pair between A and B is capped from minute 5 to minute 10 at 500\,Mbps and 100\,Mbps, respectively. All other paths remain unchanged, and the original bandwidth is restored at minute 10.
    Throughput is measured in 30 s bins and normalized by each method's top throughput during the initial healthy phase.
    \ours limits throughput loss at both bandwidth caps. Windowed \ours is additionally evaluated at 100\,Mbps, where bounded waiting further improves throughput.}
    \label{fig:slow-links-throughput}
\end{figure*}

\subsubsection{Communication Stragglers Between Islands}
\label{sec:results:communication-stragglers}

Geographically distributed training resources can form communication islands, with fast connections within sites and slower internet connections between them~\cite{jaghouar2024intellect1,lidin2026covenant72b}.
The connections between islands can also vary over time, with Internet measurements reporting substantial throughput degradation over the course of a day~\cite{mok2021measuring}.
A possible disruption is a temporary bandwidth reduction on the paths connecting two islands, while connections within each island and paths involving other islands remain available.
When several exchanges share the affected network infrastructure, the slowdown can affect multiple worker pairs, so changing peer neighborhoods can repeatedly encounter the same bottleneck.

In this setting, workers remain operational and can compute locally, but exchanges over affected paths take substantially longer, becoming communication stragglers.
A link failure is an extreme case in which no message can traverse the affected link until connectivity is restored.
Such a failure does not necessarily isolate an island, since information may still propagate through other islands when alternative paths remain available.

We emulate this scenario on the same eight-GPU setup, assigning the replicas to three logical islands containing three, three, and two replicas (\cref{fig:islands}).
Using the network namespaces described above, we reduce bandwidth in both directions between the two larger islands while leaving all other paths unchanged.
We test a cap of 500\,Mbps and a more severe cap of 100\,Mbps, applied separately to each directed worker pair between these islands rather than shared by the whole island.
Each run includes an initial healthy phase, a period of reduced bandwidth, and a final phase with the original bandwidth restored.
Workers and their control connections remain operational throughout, isolating the effect of degraded payload paths from worker failures.

We compare \ours against DDP managed by TorchFT, allowing the collective baseline to use its recovery mechanisms.
At 500 Mbps, smaller messages already limit the effect of slow links: \ours retains most of its healthy throughput, while DDP loses more than two thirds of it (\cref{fig:slow-links-500mbps}).
Further reducing the bandwidth to 100 Mbps prevents DDP from committing most of the updates during the 5-minute slowdown, while \ours retains approximately two thirds of its healthy throughput (\cref{fig:slow-links-100mbps}).
Sparsification alone allows exchanges to complete faster despite the reduced bandwidth, because each message sends less than a twentieth of the full payload.

\paragraph{Improving throughput with bounded waiting.} %
The results above wait for all requested peer contributions, but peer-to-peer exchanges allow each replica to aggregate available contributions without waiting for every requested peer.
To exploit this flexibility under communication stragglers, one could use a windowed approach to gossip, which combines background reception with a bounded communication window.
Each replica posts a receive operation in the background so that peers can send updates before it reaches its own exchange stage.
At the deadline, it aggregates only complete peer contributions and proceeds with its outer update, using only its own endpoint if none arrived in time, as it would do for coordinates that were not shared by any other peer.
A global collective instead requires the prescribed participation to complete its result, even when communication overlaps with computation.

Sparsification allows each replica to contact more peers under the same communication budget, increasing the opportunities to receive contributions before the deadline.
This benefit depends on reaching peers through unaffected paths, since peers sharing a bottleneck can be delayed together.
We therefore evaluate \ours with windowed gossip in the same island setup.
During the initial healthy period, we measure the distribution of complete sparse exchange times.
The communication window is fixed at $2\times$ the measured 99th percentile and kept unchanged during the period of reduced bandwidth to accommodate small differences in replica timing.
\Cref{fig:slow-links-100mbps} shows that windowed gossip further improves throughput under the more severe slowdown by allowing updates to proceed with fewer contributions.
With this window, committed updates used an average of $2.48 \pm 0.88$ peer contributions out of four requested peers, while retaining approximately 86\% of healthy throughput.

\section{Conclusion}
We introduced \oursbold, a decentralized training method that combines sparse parameter exchanges with error feedback to make broader peer neighborhoods affordable.
Under tight communication budgets, exchanging fewer coordinates with more peers improves training quality over denser exchanges with fewer peers.
This design can outperform dense gossip while communicating less and limiting quality degradation as the replica count grows.
Our implementation turns these smaller exchanges into higher training throughput under bandwidth constraints and maintains much of this performance during transient link slowdowns.
These results support decentralized language model pretraining over bandwidth-constrained networks.

\clearpage

\section*{Acknowledgments}
This work was supported by PEPR IA on grant SHARP ANR-23-PEIA-0008 and PEPR NUMPEX on grant  DAIMOS ANR-25-EXNU-0002.
Part of this work was granted access to the JeanZay HPC/AI resources of IDRIS under the allocation AD011015884R1, AD011017481, AD011016641, AD011017661 and AD011017766.
E.B. and P.C. acknowledge funding from NSERC Discovery and FRQNT New Scholar.

\bibliography{main}

@InProceedings{koloskova2019decentralized,
  title = 	 {Decentralized Stochastic Optimization and Gossip Algorithms with Compressed Communication},
  author =       {Koloskova, Anastasia and Stich, Sebastian and Jaggi, Martin},
  booktitle = 	 {Proceedings of the 36th International Conference on Machine Learning},
  pages = 	 {3478--3487},
  year = 	 {2019},
  editor = 	 {Chaudhuri, Kamalika and Salakhutdinov, Ruslan},
  volume = 	 {97},
  series = 	 {Proceedings of Machine Learning Research},
  month = 	 {09--15 Jun},
  publisher =    {PMLR},
  url = 	 {https://proceedings.mlr.press/v97/koloskova19a.html},
}

@inproceedings{stich2019localsgd,
title={Local {SGD} Converges Fast and Communicates Little},
author={Sebastian U. Stich},
booktitle={International Conference on Learning Representations},
year={2019},
url={https://openreview.net/forum?id=S1g2JnRcFX},
}

@inproceedings{koloskova2020unified,
  title = 	 {A Unified Theory of Decentralized {SGD} with Changing Topology and Local Updates},
  author =       {Koloskova, Anastasia and Loizou, Nicolas and Boreiri, Sadra and Jaggi, Martin and Stich, Sebastian},
  booktitle = 	 {Proceedings of the 37th International Conference on Machine Learning},
  pages = 	 {5381--5393},
  year = 	 {2020},
  editor = 	 {III, Hal Daumé and Singh, Aarti},
  volume = 	 {119},
  series = 	 {Proceedings of Machine Learning Research},
  month = 	 {13--18 Jul},
  publisher =    {PMLR},
  url = 	 {https://proceedings.mlr.press/v119/koloskova20a.html},
}

@article{li2020pytorchdistributed,
  author = {Li, Shen and Zhao, Yanli and Varma, Rohan and Salpekar, Omkar and Noordhuis, Pieter and Li, Teng and Paszke, Adam and Smith, Jeff and Vaughan, Brian and Damania, Pritam and Chintala, Soumith},
  title = {PyTorch distributed: experiences on accelerating data parallel training},
  year = {2020},
  publisher = {VLDB Endowment},
  volume = {13},
  number = {12},
  journal = {Proc. VLDB Endow.},
}

@article{patarasuk2009bandwidth,
  author = {Patarasuk, Pitch and Yuan, Xin},
  title = {Bandwidth optimal all-reduce algorithms for clusters of workstations},
  year = {2009},
  issue_date = {February, 2009},
  publisher = {Academic Press, Inc.},
  address = {USA},
  volume = {69},
  number = {2},
  journal = {J. Parallel Distrib. Comput.},
  pages = {117–124},
  numpages = {8},
}

@inproceedings{mok2021measuring,
  author = {Mok, Ricky K. P. and Zou, Hongyu and Yang, Rui and Koch, Tom and Katz-Bassett, Ethan and Claffy, Kimberly C.},
  title = {Measuring the Network Performance of {Google Cloud Platform}},
  booktitle = {Proceedings of the 21st ACM Internet Measurement Conference},
  year = {2021},
  pages = {54--61},
  doi = {10.1145/3487552.3487862},
  url = {https://doi.org/10.1145/3487552.3487862},
}

@misc{jaghouar2024intellect1,
  title = {INTELLECT-1 Technical Report},
  author = {Sami Jaghouar and Jack Min Ong and Manveer Basra and Fares Obeid and Jannik Straube and Michael Keiblinger and Elie Bakouch and Lucas Atkins and Maziyar Panahi and Charles Goddard and Max Ryabinin and Johannes Hagemann},
  year = {2024},
  eprint = {2412.01152},
  archiveprefix = {arXiv},
  primaryclass = {cs.DC},
  url = {https://arxiv.org/abs/2412.01152},
}

@misc{lidin2026covenant72b,
  title = {Covenant-72B: Pre-Training a 72B LLM with Trustless Peers Over-the-Internet},
  author = {Joel Lidin and Amir Sarfi and Erfan Miahi and Quentin Anthony and Shivam Chauhan and Evangelos Pappas and Benjamin Thérien and Eugene Belilovsky and Samuel Dare},
  year = {2026},
  eprint = {2603.08163},
  archiveprefix = {arXiv},
  primaryclass = {cs.DC},
  url = {https://arxiv.org/abs/2603.08163},
}

@misc{cagnasso2026unifying,
  title = {Unifying Local Communications and Local Updates for LLM Pretraining},
  author = {Pietro Cagnasso and Eugene Belilovsky and Edouard Oyallon},
  year = {2026},
  eprint = {2606.11081},
  archiveprefix = {arXiv},
  primaryclass = {cs.LG},
  url = {https://arxiv.org/abs/2606.11081},
}

@misc{horoi2026model,
  title = {Can Model Merging Improve Aggregation in DiLoCo?},
  author = {Stefan Horoi and Benjamin Thérien and Guy Wolf and Eugene Belilovsky},
  year = {2026},
  eprint = {2607.03011},
  archiveprefix = {arXiv},
  primaryclass = {cs.LG},
  url = {https://arxiv.org/abs/2607.03011},
}

@article{nabli2025decentralized,
  title={Decentralized asynchronous optimization with dadao allows decoupling and acceleration},
  author={Nabli, Adel and Oyallon, Edouard},
  journal={Journal of Machine Learning Research},
  volume={26},
  number={207},
  pages={1--48},
  year={2025}
}

@misc{
kopyshvskiy2026lacco,
title={{LACCO}: Local Accumulation with Computation{\textendash}Communication Overlap},
author={Daniele Kopyshevskiy and Benjamin Thérien and Nicolas Pinto and Edouard Oyallon},
year={2026},
url={https://arxiv.org/abs/2611.17991},
}

@inproceedings{
nabli2024acco,
title={{ACCO}: Accumulate while you Communicate, Hiding Communications in Distributed {LLM} Training},
author={Adel Nabli and Louis Fournier and Pierre ERBACHER and Louis Serrano and Eugene Belilovsky and Edouard Oyallon},
booktitle={OPT 2024: Optimization for Machine Learning},
year={2024},
url={https://openreview.net/forum?id=VrKAGrUqGe}
}

@inproceedings{nabli2023dadao,
  title={DADAO: Decoupled accelerated decentralized asynchronous optimization},
  author={Nabli, Adel and Oyallon, Edouard},
  booktitle={International Conference on Machine Learning},
  pages={25604--25626},
  year={2023},
  organization={PMLR}
}

@misc{belilovsky2026how,
  title = {How Inner Optimizers Shape Replica Scaling},
  author = {Eugene Belilovsky and Stefan Horoi and Edouard Oyallon and Benjamin Th{\'e}rien},
  year = {2026},
  eprint = {2610.19064},
  archiveprefix = {arXiv},
  primaryclass = {cs.LG},
  url = {https://arxiv.org/abs/2610.19064},
}

@inproceedings{douillard2023diloco,
  title = {DiLoCo: Distributed Low-Communication Training of Language Models},
  author = {Arthur Douillard and Qixuan Feng and Andrei Alex Rusu and Rachita Chhaparia and Yani Donchev and Adhiguna Kuncoro and MarcAurelio Ranzato and Arthur Szlam and Jiajun Shen},
  booktitle = {2nd Workshop on Advancing Neural Network Training: Computational Efficiency, Scalability, and Resource Optimization (WANT@ICML 2024)},
  year = {2024},
  url = {https://openreview.net/forum?id=pICSfWkJIk},
}

@misc{sarfi2026sparseloco,
      title={Overcoming the Communication-Performance Tradeoff in LLM Pretraining}, 
      author={Amir Sarfi and Benjamin Thérien and Joel Lidin and Eugene Belilovsky},
      year={2026},
      eprint={2508.15706},
      archivePrefix={arXiv},
      primaryClass={cs.LG},
      url={https://arxiv.org/abs/2508.15706}, 
}

@inproceedings{wang2023cocktailsgd,
  title = {{C}ocktail{SGD}: Fine-tuning Foundation Models over 500{M}bps Networks},
  author = {Wang, Jue and Lu, Yucheng and Yuan, Binhang and Chen, Beidi and Liang, Percy and De Sa, Christopher and Re, Christopher and Zhang, Ce},
  booktitle = {Proceedings of the 40th International Conference on Machine Learning},
  pages = {36058--36076},
  year = {2023},
  editor = {Krause, Andreas and Brunskill, Emma and Cho, Kyunghyun and Engelhardt, Barbara and Sabato, Sivan and Scarlett, Jonathan},
  volume = {202},
  series = {Proceedings of Machine Learning Research},
  month = {23--29 Jul},
  publisher = {PMLR},
  url = {https://proceedings.mlr.press/v202/wang23t.html},
}

@article{stich2021error,
  author  = {Sebastian U. Stich and Sai Praneeth Karimireddy},
  title   = {The Error-Feedback framework: SGD with Delayed Gradients},
  journal = {Journal of Machine Learning Research},
  year    = {2020},
  volume  = {21},
  number  = {237},
  pages   = {1--36},
  url     = {http://jmlr.org/papers/v21/19-748.html}
}

@article{boyd2006randomized,
  title = {Randomized gossip algorithms},
  author = {Boyd, Stephen and Ghosh, Arpita and Prabhakar, Balaji and Shah, Devavrat},
  journal = {IEEE transactions on information theory},
  volume = {52},
  number = {6},
  pages = {2508--2530},
  year = {2006},
  publisher = {IEEE},
}

@inproceedings{lian2017can,
  author = {Lian, Xiangru and Zhang, Ce and Zhang, Huan and Hsieh, Cho-Jui and Zhang, Wei and Liu, Ji},
  title = {Can decentralized algorithms outperform centralized algorithms? a case study for decentralized parallel stochastic gradient descent},
  year = {2017},
  publisher = {Curran Associates Inc.},
  booktitle = {Proceedings of the 31st International Conference on Neural Information Processing Systems},
  pages = {5336–5346},
  numpages = {11},
  location = {Long Beach, California, USA},
  series = {NIPS'17},
}

@inproceedings{assran2019stochastic,
  title = {Stochastic gradient push for distributed deep learning},
  author = {Assran, Mahmoud and Loizou, Nicolas and Ballas, Nicolas and Rabbat, Mike},
  booktitle = {International Conference on Machine Learning},
  pages = {344--353},
  year = {2019},
  organization = {PMLR},
}

@inproceedings{wang2024promise,
  author = {Wang, Zesen and Zhang, Jiaojiao and Wu, Xuyang and Johansson, Mikael},
  booktitle = {International Conference on Learning Representations},
  editor = {Y. Yue and A. Garg and N. Peng and F. Sha and R. Yu},
  pages = {26204--26249},
  title = {From Promise to Practice: Realizing High-performance Decentralized Training},
  volume = {2025},
  year = {2025},
}

@inproceedings{mcmahan2017communication,
  title = {{Communication-Efficient Learning of Deep Networks from Decentralized Data}},
  author = {McMahan, Brendan and Moore, Eider and Ramage, Daniel and Hampson, Seth and Arcas, Blaise Aguera y},
  booktitle = {Proceedings of the 20th International Conference on Artificial Intelligence and Statistics},
  pages = {1273--1282},
  year = {2017},
  editor = {Singh, Aarti and Zhu, Jerry},
  volume = {54},
  series = {Proceedings of Machine Learning Research},
  month = {20--22 Apr},
  publisher = {PMLR},
  url = {https://proceedings.mlr.press/v54/mcmahan17a.html},
}

@inproceedings{reddi2021adaptive,
  title = {Adaptive Federated Optimization},
  author = {Sashank J. Reddi and Zachary Charles and Manzil Zaheer and Zachary Garrett and Keith Rush and Jakub Kone{\v{c}}n{\'y} and Sanjiv Kumar and Hugh Brendan McMahan},
  booktitle = {International Conference on Learning Representations},
  year = {2021},
  url = {https://openreview.net/forum?id=LkFG3lB13U5},
}

@inproceedings{stich2018sparsified,
 author = {Stich, Sebastian U and Cordonnier, Jean-Baptiste and Jaggi, Martin},
 booktitle = {Advances in Neural Information Processing Systems},
 editor = {S. Bengio and H. Wallach and H. Larochelle and K. Grauman and N. Cesa-Bianchi and R. Garnett},
 pages = {},
 publisher = {Curran Associates, Inc.},
 title = {Sparsified SGD with Memory},
 url = {https://proceedings.neurips.cc/paper_files/paper/2018/file/b440509a0106086a67bc2ea9df0a1dab-Paper.pdf},
 volume = {31},
 year = {2018}
}

@InProceedings{karimireddy2019error,
  title = 	 {Error Feedback Fixes {S}ign{SGD} and other Gradient Compression Schemes},
  author =       {Karimireddy, Sai Praneeth and Rebjock, Quentin and Stich, Sebastian and Jaggi, Martin},
  booktitle = 	 {Proceedings of the 36th International Conference on Machine Learning},
  pages = 	 {3252--3261},
  year = 	 {2019},
  editor = 	 {Chaudhuri, Kamalika and Salakhutdinov, Ruslan},
  volume = 	 {97},
  series = 	 {Proceedings of Machine Learning Research},
  month = 	 {09--15 Jun},
  publisher =    {PMLR},
  url = 	 {https://proceedings.mlr.press/v97/karimireddy19a.html},
}

@inproceedings{koloskova2020decentralized,
title={Decentralized Deep Learning with Arbitrary Communication Compression},
author={Anastasia Koloskova* and Tao Lin* and Sebastian U Stich and Martin Jaggi},
booktitle={International Conference on Learning Representations},
year={2020}
}

@inproceedings{
peng2024demo,
title={DeMo: Decoupled Momentum Optimization},
author={Bowen Peng and Lizhang Chen and Baiyu Su and Jeffrey Quesnelle and Diederik P Kingma and qiang liu},
booktitle={The Fourteenth International Conference on Learning Representations},
year={2026},
url={https://openreview.net/forum?id=U9oewpa7cn}
}

@INPROCEEDINGS{wang2019matcha,
  author={Wang, Jianyu and Sahu, Anit Kumar and Yang, Zhouyi and Joshi, Gauri and Kar, Soummya},
  booktitle={2019 Sixth Indian Control Conference (ICC)}, 
  title={MATCHA: Speeding Up Decentralized SGD via Matching Decomposition Sampling}, 
  year={2019},
  volume={},
  number={},
  pages={299-300}
  }

@inproceedings{song2022communication,
title={Communication-Efficient Topologies for Decentralized Learning with \$O(1)\$ Consensus Rate},
author={Zhuoqing Song and Weijian Li and Kexin Jin and Lei Shi and Ming Yan and Wotao Yin and Kun Yuan},
booktitle={Advances in Neural Information Processing Systems},
editor={Alice H. Oh and Alekh Agarwal and Danielle Belgrave and Kyunghyun Cho},
year={2022},
url={https://openreview.net/forum?id=AyiiHcRzTd}
}

@article{vogels2023beyond,
author = {Vogels, Thijs and Hendrikx, Hadrien and Jaggi, Martin},
title = {Beyond spectral gap: the role of the topology in decentralized learning},
year = {2023},
issue_date = {January 2023},
publisher = {JMLR.org},
volume = {24},
number = {1},
issn = {1532-4435},
journal = {J. Mach. Learn. Res.},
month = jan,
articleno = {355},
numpages = {31}
}

@misc{deepseekai2025deepseekv3,
      title={DeepSeek-V3 Technical Report}, 
      author={DeepSeek-AI and Aixin Liu and Bei Feng and Bing Xue and Bingxuan Wang and Bochao Wu and Chengda Lu and Chenggang Zhao and Chengqi Deng and Chenyu Zhang and Chong Ruan and Damai Dai and Daya Guo and Dejian Yang and Deli Chen and Dongjie Ji and Erhang Li and Fangyun Lin and Fucong Dai and Fuli Luo and Guangbo Hao and Guanting Chen and Guowei Li and H. Zhang and Han Bao and Hanwei Xu and Haocheng Wang and Haowei Zhang and Honghui Ding and Huajian Xin and Huazuo Gao and Hui Li and Hui Qu and J. L. Cai and Jian Liang and Jianzhong Guo and Jiaqi Ni and Jiashi Li and Jiawei Wang and Jin Chen and Jingchang Chen and Jingyang Yuan and Junjie Qiu and Junlong Li and Junxiao Song and Kai Dong and Kai Hu and Kaige Gao and Kang Guan and Kexin Huang and Kuai Yu and Lean Wang and Lecong Zhang and Lei Xu and Leyi Xia and Liang Zhao and Litong Wang and Liyue Zhang and Meng Li and Miaojun Wang and Mingchuan Zhang and Minghua Zhang and Minghui Tang and Mingming Li and Ning Tian and Panpan Huang and Peiyi Wang and Peng Zhang and Qiancheng Wang and Qihao Zhu and Qinyu Chen and Qiushi Du and R. J. Chen and R. L. Jin and Ruiqi Ge and Ruisong Zhang and Ruizhe Pan and Runji Wang and Runxin Xu and Ruoyu Zhang and Ruyi Chen and S. S. Li and Shanghao Lu and Shangyan Zhou and Shanhuang Chen and Shaoqing Wu and Shengfeng Ye and Shengfeng Ye and Shirong Ma and Shiyu Wang and Shuang Zhou and Shuiping Yu and Shunfeng Zhou and Shuting Pan and T. Wang and Tao Yun and Tian Pei and Tianyu Sun and W. L. Xiao and Wangding Zeng and Wanjia Zhao and Wei An and Wen Liu and Wenfeng Liang and Wenjun Gao and Wenqin Yu and Wentao Zhang and X. Q. Li and Xiangyue Jin and Xianzu Wang and Xiao Bi and Xiaodong Liu and Xiaohan Wang and Xiaojin Shen and Xiaokang Chen and Xiaokang Zhang and Xiaosha Chen and Xiaotao Nie and Xiaowen Sun and Xiaoxiang Wang and Xin Cheng and Xin Liu and Xin Xie and Xingchao Liu and Xingkai Yu and Xinnan Song and Xinxia Shan and Xinyi Zhou and Xinyu Yang and Xinyuan Li and Xuecheng Su and Xuheng Lin and Y. K. Li and Y. Q. Wang and Y. X. Wei and Y. X. Zhu and Yang Zhang and Yanhong Xu and Yanhong Xu and Yanping Huang and Yao Li and Yao Zhao and Yaofeng Sun and Yaohui Li and Yaohui Wang and Yi Yu and Yi Zheng and Yichao Zhang and Yifan Shi and Yiliang Xiong and Ying He and Ying Tang and Yishi Piao and Yisong Wang and Yixuan Tan and Yiyang Ma and Yiyuan Liu and Yongqiang Guo and Yu Wu and Yuan Ou and Yuchen Zhu and Yuduan Wang and Yue Gong and Yuheng Zou and Yujia He and Yukun Zha and Yunfan Xiong and Yunxian Ma and Yuting Yan and Yuxiang Luo and Yuxiang You and Yuxuan Liu and Yuyang Zhou and Z. F. Wu and Z. Z. Ren and Zehui Ren and Zhangli Sha and Zhe Fu and Zhean Xu and Zhen Huang and Zhen Zhang and Zhenda Xie and Zhengyan Zhang and Zhewen Hao and Zhibin Gou and Zhicheng Ma and Zhigang Yan and Zhihong Shao and Zhipeng Xu and Zhiyu Wu and Zhongyu Zhang and Zhuoshu Li and Zihui Gu and Zijia Zhu and Zijun Liu and Zilin Li and Ziwei Xie and Ziyang Song and Ziyi Gao and Zizheng Pan},
      year={2025},
      eprint={2412.19437},
      archivePrefix={arXiv},
      primaryClass={cs.CL},
      url={https://arxiv.org/abs/2412.19437}, 
}

@inproceedings{ying2021exponential,
 author = {Ying, Bicheng and Yuan, Kun and Chen, Yiming and Hu, Hanbin and PAN, PAN and Yin, Wotao},
 booktitle = {Advances in Neural Information Processing Systems},
 editor = {M. Ranzato and A. Beygelzimer and Y. Dauphin and P.S. Liang and J. Wortman Vaughan},
 pages = {13975--13987},
 publisher = {Curran Associates, Inc.},
 title = {Exponential Graph is Provably Efficient for Decentralized Deep Training},
 url = {https://proceedings.neurips.cc/paper_files/paper/2021/file/74e1ed8b55ea44fd7dbb685c412568a4-Paper.pdf},
 volume = {34},
 year = {2021}
}

@inproceedings{liang2025torchtitan,
  title = {TorchTitan: One-stop PyTorch native solution for production ready {LLM} pretraining},
  author = {Wanchao Liang and Tianyu Liu and Less Wright and Will Constable and Andrew Gu and Chien-Chin Huang and Iris Zhang and Wei Feng and Howard Huang and Junjie Wang and Sanket Purandare and Gokul Nadathur and Stratos Idreos},
  booktitle = {The Thirteenth International Conference on Learning Representations},
  year = {2025},
  url = {https://openreview.net/forum?id=SFN6Wm7YBI},
}

@article{touvron2023llama2,
  title = {Llama 2: Open Foundation and Fine-Tuned Chat Models},
  author = {Hugo Touvron and Louis Martin and Kevin Stone and Peter Albert and Amjad Almahairi and Yasmine Babaei and Nikolay Bashlykov and Soumya Batra and Prajjwal Bhargava and Shruti Bhosale and Dan Bikel and Lukas Blecher and Cristian Canton Ferrer and Moya Chen and Guillem Cucurull and David Esiobu and Jude Fernandes and Jeremy Fu and Wenyin Fu and Brian Fuller and Cynthia Gao and Vedanuj Goswami and Naman Goyal and Anthony Hartshorn and Saghar Hosseini and Rui Hou and Hakan Inan and Marcin Kardas and Viktor Kerkez and Madian Khabsa and Isabel Kloumann and Artem Korenev and Punit Singh Koura and Marie-Anne Lachaux and Thibaut Lavril and Jenya Lee and Diana Liskovich and Yinghai Lu and Yuning Mao and Xavier Martinet and Todor Mihaylov and Pushkar Mishra and Igor Molybog and Yixin Nie and Andrew Poulton and Jeremy Reizenstein and Rashi Rungta and Kalyan Saladi and Alan Schelten and Ruan Silva and Eric Michael Smith and Ranjan Subramanian and Xiaoqing Ellen Tan and Binh Tang and Ross Taylor and Adina Williams and Jian Xiang Kuan and Puxin Xu and Zheng Yan and Iliyan Zarov and Yuchen Zhang and Angela Fan and Melanie Kambadur and Sharan Narang and Aurelien Rodriguez and Robert Stojnic and Sergey Edunov and Thomas Scialom},
  year = {2023},
  journal = {arXiv preprint arXiv:2307.09288},
}

@misc{grattafiori2024llama,
      title={The Llama 3 Herd of Models}, 
      author={Aaron Grattafiori and Abhimanyu Dubey and Abhinav Jauhri and Abhinav Pandey and Abhishek Kadian and Ahmad Al-Dahle and Aiesha Letman and Akhil Mathur and Alan Schelten and Alex Vaughan and Amy Yang and Angela Fan and Anirudh Goyal and Anthony Hartshorn and Aobo Yang and Archi Mitra and Archie Sravankumar and Artem Korenev and Arthur Hinsvark and Arun Rao and Aston Zhang and Aurelien Rodriguez and Austen Gregerson and Ava Spataru and Baptiste Roziere and Bethany Biron and Binh Tang and Bobbie Chern and Charlotte Caucheteux and Chaya Nayak and Chloe Bi and Chris Marra and Chris McConnell and Christian Keller and Christophe Touret and Chunyang Wu and Corinne Wong and Cristian Canton Ferrer and Cyrus Nikolaidis and Damien Allonsius and Daniel Song and Danielle Pintz and Danny Livshits and Danny Wyatt and David Esiobu and Dhruv Choudhary and Dhruv Mahajan and Diego Garcia-Olano and Diego Perino and Dieuwke Hupkes and Egor Lakomkin and Ehab AlBadawy and Elina Lobanova and Emily Dinan and Eric Michael Smith and Filip Radenovic and Francisco Guzmán and Frank Zhang and Gabriel Synnaeve and Gabrielle Lee and Georgia Lewis Anderson and Govind Thattai and Graeme Nail and Gregoire Mialon and Guan Pang and Guillem Cucurell and Hailey Nguyen and Hannah Korevaar and Hu Xu and Hugo Touvron and Iliyan Zarov and Imanol Arrieta Ibarra and Isabel Kloumann and Ishan Misra and Ivan Evtimov and Jack Zhang and Jade Copet and Jaewon Lee and Jan Geffert and Jana Vranes and Jason Park and Jay Mahadeokar and Jeet Shah and Jelmer van der Linde and Jennifer Billock and Jenny Hong and Jenya Lee and Jeremy Fu and Jianfeng Chi and Jianyu Huang and Jiawen Liu and Jie Wang and Jiecao Yu and Joanna Bitton and Joe Spisak and Jongsoo Park and Joseph Rocca and Joshua Johnstun and Joshua Saxe and Junteng Jia and Kalyan Vasuden Alwala and Karthik Prasad and Kartikeya Upasani and Kate Plawiak and Ke Li and Kenneth Heafield and Kevin Stone and Khalid El-Arini and Krithika Iyer and Kshitiz Malik and Kuenley Chiu and Kunal Bhalla and Kushal Lakhotia and Lauren Rantala-Yeary and Laurens van der Maaten and Lawrence Chen and Liang Tan and Liz Jenkins and Louis Martin and Lovish Madaan and Lubo Malo and Lukas Blecher and Lukas Landzaat and Luke de Oliveira and Madeline Muzzi and Mahesh Pasupuleti and Mannat Singh and Manohar Paluri and Marcin Kardas and Maria Tsimpoukelli and Mathew Oldham and Mathieu Rita and Maya Pavlova and Melanie Kambadur and Mike Lewis and Min Si and Mitesh Kumar Singh and Mona Hassan and Naman Goyal and Narjes Torabi and Nikolay Bashlykov and Nikolay Bogoychev and Niladri Chatterji and Ning Zhang and Olivier Duchenne and Onur Çelebi and Patrick Alrassy and Pengchuan Zhang and Pengwei Li and Petar Vasic and Peter Weng and Prajjwal Bhargava and Pratik Dubal and Praveen Krishnan and Punit Singh Koura and Puxin Xu and Qing He and Qingxiao Dong and Ragavan Srinivasan and Raj Ganapathy and Ramon Calderer and Ricardo Silveira Cabral and Robert Stojnic and Roberta Raileanu and Rohan Maheswari and Rohit Girdhar and Rohit Patel and Romain Sauvestre and Ronnie Polidoro and Roshan Sumbaly and Ross Taylor and Ruan Silva and Rui Hou and Rui Wang and Saghar Hosseini and Sahana Chennabasappa and Sanjay Singh and Sean Bell and Seohyun Sonia Kim and Sergey Edunov and Shaoliang Nie and Sharan Narang and Sharath Raparthy and Sheng Shen and Shengye Wan and Shruti Bhosale and Shun Zhang and Simon Vandenhende and Soumya Batra and Spencer Whitman and Sten Sootla and Stephane Collot and Suchin Gururangan and Sydney Borodinsky and Tamar Herman and Tara Fowler and Tarek Sheasha and Thomas Georgiou and Thomas Scialom and Tobias Speckbacher and Todor Mihaylov and Tong Xiao and Ujjwal Karn and Vedanuj Goswami and Vibhor Gupta and Vignesh Ramanathan and Viktor Kerkez and Vincent Gonguet and Virginie Do and Vish Vogeti and Vítor Albiero and Vladan Petrovic and Weiwei Chu and Wenhan Xiong and Wenyin Fu and Whitney Meers and Xavier Martinet and Xiaodong Wang and Xiaofang Wang and Xiaoqing Ellen Tan and Xide Xia and Xinfeng Xie and Xuchao Jia and Xuewei Wang and Yaelle Goldschlag and Yashesh Gaur and Yasmine Babaei and Yi Wen and Yiwen Song and Yuchen Zhang and Yue Li and Yuning Mao and Zacharie Delpierre Coudert and Zheng Yan and Zhengxing Chen and Zoe Papakipos and Aaditya Singh and Aayushi Srivastava and Abha Jain and Adam Kelsey and Adam Shajnfeld and Adithya Gangidi and Adolfo Victoria and Ahuva Goldstand and Ajay Menon and Ajay Sharma and Alex Boesenberg and Alexei Baevski and Allie Feinstein and Amanda Kallet and Amit Sangani and Amos Teo and Anam Yunus and Andrei Lupu and Andres Alvarado and Andrew Caples and Andrew Gu and Andrew Ho and Andrew Poulton and Andrew Ryan and Ankit Ramchandani and Annie Dong and Annie Franco and Anuj Goyal and Aparajita Saraf and Arkabandhu Chowdhury and Ashley Gabriel and Ashwin Bharambe and Assaf Eisenman and Azadeh Yazdan and Beau James and Ben Maurer and Benjamin Leonhardi and Bernie Huang and Beth Loyd and Beto De Paola and Bhargavi Paranjape and Bing Liu and Bo Wu and Boyu Ni and Braden Hancock and Bram Wasti and Brandon Spence and Brani Stojkovic and Brian Gamido and Britt Montalvo and Carl Parker and Carly Burton and Catalina Mejia and Ce Liu and Changhan Wang and Changkyu Kim and Chao Zhou and Chester Hu and Ching-Hsiang Chu and Chris Cai and Chris Tindal and Christoph Feichtenhofer and Cynthia Gao and Damon Civin and Dana Beaty and Daniel Kreymer and Daniel Li and David Adkins and David Xu and Davide Testuggine and Delia David and Devi Parikh and Diana Liskovich and Didem Foss and Dingkang Wang and Duc Le and Dustin Holland and Edward Dowling and Eissa Jamil and Elaine Montgomery and Eleonora Presani and Emily Hahn and Emily Wood and Eric-Tuan Le and Erik Brinkman and Esteban Arcaute and Evan Dunbar and Evan Smothers and Fei Sun and Felix Kreuk and Feng Tian and Filippos Kokkinos and Firat Ozgenel and Francesco Caggioni and Frank Kanayet and Frank Seide and Gabriela Medina Florez and Gabriella Schwarz and Gada Badeer and Georgia Swee and Gil Halpern and Grant Herman and Grigory Sizov and Guangyi and Zhang and Guna Lakshminarayanan and Hakan Inan and Hamid Shojanazeri and Han Zou and Hannah Wang and Hanwen Zha and Haroun Habeeb and Harrison Rudolph and Helen Suk and Henry Aspegren and Hunter Goldman and Hongyuan Zhan and Ibrahim Damlaj and Igor Molybog and Igor Tufanov and Ilias Leontiadis and Irina-Elena Veliche and Itai Gat and Jake Weissman and James Geboski and James Kohli and Janice Lam and Japhet Asher and Jean-Baptiste Gaya and Jeff Marcus and Jeff Tang and Jennifer Chan and Jenny Zhen and Jeremy Reizenstein and Jeremy Teboul and Jessica Zhong and Jian Jin and Jingyi Yang and Joe Cummings and Jon Carvill and Jon Shepard and Jonathan McPhie and Jonathan Torres and Josh Ginsburg and Junjie Wang and Kai Wu and Kam Hou U and Karan Saxena and Kartikay Khandelwal and Katayoun Zand and Kathy Matosich and Kaushik Veeraraghavan and Kelly Michelena and Keqian Li and Kiran Jagadeesh and Kun Huang and Kunal Chawla and Kyle Huang and Lailin Chen and Lakshya Garg and Lavender A and Leandro Silva and Lee Bell and Lei Zhang and Liangpeng Guo and Licheng Yu and Liron Moshkovich and Luca Wehrstedt and Madian Khabsa and Manav Avalani and Manish Bhatt and Martynas Mankus and Matan Hasson and Matthew Lennie and Matthias Reso and Maxim Groshev and Maxim Naumov and Maya Lathi and Meghan Keneally and Miao Liu and Michael L. Seltzer and Michal Valko and Michelle Restrepo and Mihir Patel and Mik Vyatskov and Mikayel Samvelyan and Mike Clark and Mike Macey and Mike Wang and Miquel Jubert Hermoso and Mo Metanat and Mohammad Rastegari and Munish Bansal and Nandhini Santhanam and Natascha Parks and Natasha White and Navyata Bawa and Nayan Singhal and Nick Egebo and Nicolas Usunier and Nikhil Mehta and Nikolay Pavlovich Laptev and Ning Dong and Norman Cheng and Oleg Chernoguz and Olivia Hart and Omkar Salpekar and Ozlem Kalinli and Parkin Kent and Parth Parekh and Paul Saab and Pavan Balaji and Pedro Rittner and Philip Bontrager and Pierre Roux and Piotr Dollar and Polina Zvyagina and Prashant Ratanchandani and Pritish Yuvraj and Qian Liang and Rachad Alao and Rachel Rodriguez and Rafi Ayub and Raghotham Murthy and Raghu Nayani and Rahul Mitra and Rangaprabhu Parthasarathy and Raymond Li and Rebekkah Hogan and Robin Battey and Rocky Wang and Russ Howes and Ruty Rinott and Sachin Mehta and Sachin Siby and Sai Jayesh Bondu and Samyak Datta and Sara Chugh and Sara Hunt and Sargun Dhillon and Sasha Sidorov and Satadru Pan and Saurabh Mahajan and Saurabh Verma and Seiji Yamamoto and Sharadh Ramaswamy and Shaun Lindsay and Shaun Lindsay and Sheng Feng and Shenghao Lin and Shengxin Cindy Zha and Shishir Patil and Shiva Shankar and Shuqiang Zhang and Shuqiang Zhang and Sinong Wang and Sneha Agarwal and Soji Sajuyigbe and Soumith Chintala and Stephanie Max and Stephen Chen and Steve Kehoe and Steve Satterfield and Sudarshan Govindaprasad and Sumit Gupta and Summer Deng and Sungmin Cho and Sunny Virk and Suraj Subramanian and Sy Choudhury and Sydney Goldman and Tal Remez and Tamar Glaser and Tamara Best and Thilo Koehler and Thomas Robinson and Tianhe Li and Tianjun Zhang and Tim Matthews and Timothy Chou and Tzook Shaked and Varun Vontimitta and Victoria Ajayi and Victoria Montanez and Vijai Mohan and Vinay Satish Kumar and Vishal Mangla and Vlad Ionescu and Vlad Poenaru and Vlad Tiberiu Mihailescu and Vladimir Ivanov and Wei Li and Wenchen Wang and Wenwen Jiang and Wes Bouaziz and Will Constable and Xiaocheng Tang and Xiaojian Wu and Xiaolan Wang and Xilun Wu and Xinbo Gao and Yaniv Kleinman and Yanjun Chen and Ye Hu and Ye Jia and Ye Qi and Yenda Li and Yilin Zhang and Ying Zhang and Yossi Adi and Youngjin Nam and Yu and Wang and Yu Zhao and Yuchen Hao and Yundi Qian and Yunlu Li and Yuzi He and Zach Rait and Zachary DeVito and Zef Rosnbrick and Zhaoduo Wen and Zhenyu Yang and Zhiwei Zhao and Zhiyu Ma},
      year={2024},
      eprint={2407.21783},
      archivePrefix={arXiv},
      primaryClass={cs.AI},
      url={https://arxiv.org/abs/2407.21783}, 
}

@inproceedings{muennighoff2025olmoe,
  title = {{OLMoE}: Open Mixture-of-Experts Language Models},
  author = {Muennighoff, Niklas and Soldaini, Luca and Groeneveld, Dirk and Lo, Kyle and Morrison, Jacob and Min, Sewon and Shi, Weijia and Walsh, Pete and Tafjord, Oyvind and Lambert, Nathan and Gu, Yuling and Arora, Shane and Bhagia, Akshita and Schwenk, Dustin and Wadden, David and Wettig, Alexander and Hui, Binyuan and Dettmers, Tim and Kiela, Douwe and Farhadi, Ali and Smith, Noah A. and Koh, Pang Wei and Singh, Amanpreet and Hajishirzi, Hannaneh},
  booktitle = {International Conference on Learning Representations},
  year = {2025},
  url = {https://openreview.net/forum?id=xXTkbTBmqq},
}

@inproceedings{penedo2024fineweb,
  title = {The FineWeb Datasets: Decanting the Web for the Finest Text Data at Scale},
  author = {Guilherme Penedo and Hynek Kydl{\'\i}{\v{c}}ek and Loubna Ben allal and Anton Lozhkov and Margaret Mitchell and Colin Raffel and Leandro Von Werra and Thomas Wolf},
  booktitle = {The Thirty-eighth Conference on Neural Information Processing Systems Datasets and Benchmarks Track},
  year = {2024},
  url = {https://openreview.net/forum?id=n6SCkn2QaG},
}

@inproceedings{hoffmann2022training,
  title={Training Compute-Optimal Large Language Models}, 
  author = {Hoffmann, Jordan and Borgeaud, Sebastian and Mensch, Arthur and Buchatskaya, Elena and Cai, Trevor and Rutherford, Eliza and de Las Casas, Diego and Hendricks, Lisa Anne and Welbl, Johannes and Clark, Aidan and Hennigan, Thomas and Noland, Eric and Millican, Katherine and van den Driessche, George and Damoc, Bogdan and Guy, Aurelia and Osindero, Simon and Simonyan, Kar\'{e}n and Elsen, Erich and Vinyals, Oriol and Rae, Jack and Sifre, Laurent},
  booktitle = {Advances in Neural Information Processing Systems},
  volume = {35},
  year = {2022},
}

@inproceedings{nabli2023acid,
 author = {Nabli, Adel and Belilovsky, Eugene and Oyallon, Edouard},
 booktitle = {Advances in Neural Information Processing Systems},
 doi = {10.52202/075280-2054},
 editor = {A. Oh and T. Naumann and A. Globerson and K. Saenko and M. Hardt and S. Levine},
 pages = {47451--47474},
 publisher = {Curran Associates, Inc.},
 title = {$\mathrm{A}^2\mathrm{CiD}^2$: Accelerating Asynchronous Communication in Decentralized Deep Learning},
 url={https://openreview.net/forum?id=YE04aRkeZb},
 volume = {36},
 year = {2023}
}

@inproceedings{
fournier2024wash,
title={{WASH}: Train your Ensemble with Communication-Efficient Weight Shuffling, then Average},
author={Louis Fournier and Adel Nabli and Masih Aminbeidokhti and Marco Pedersoli and Eugene Belilovsky and Edouard Oyallon},
booktitle={OPT 2024: Optimization for Machine Learning},
year={2024},
url={https://openreview.net/forum?id=l3gJlQB8oc}
}

@inproceedings{fatkhullin2023momentum,
  title = {Momentum Provably Improves Error Feedback!},
  author = {Ilyas Fatkhullin and Alexander Tyurin and Peter Richt{\'{a}}rik},
  booktitle = {Advances in Neural Information Processing Systems},
  volume = {36},
  year = {2023},
}

@inproceedings{basu2019qsparse,
title = {Qsparse-local-SGD: distributed SGD with quantization, sparsification, and local computations},
author={Debraj Basu and Deepesh Data and Can Karakus and Suhas Diggavi},
year = {2019},
publisher = {Curran Associates Inc.},
address = {Red Hook, NY, USA},
booktitle = {Proceedings of the 33rd International Conference on Neural Information Processing Systems},
articleno = {1316},
numpages = {12}
}

@misc{kolehmainen2026noloco,
      title={NoLoCo: No-all-reduce Low Communication Training Method for Large Models}, 
      author={Jari Kolehmainen and Nikolay Blagoev and Semih Kara and John Donaghy and Christopher Nies and Oğuzhan Ersoy},
      year={2026},
      eprint={2506.10911},
      archivePrefix={arXiv},
      primaryClass={cs.LG},
      url={https://arxiv.org/abs/2506.10911}, 
}

@inproceedings{loeschcke2024loqt,
author = {Loeschcke, Sebastian and Toftrup, Mads and Kastoryano, Michael J. and Belongie, Serge and Sn{\ae}bjarnarson, V{\'e}steinn},
title = {LoQT: low-rank adapters for quantized pretraining},
year = {2024},
booktitle = {Proceedings of the 38th International Conference on Neural Information Processing Systems},
articleno = {3661},
}

@inproceedings{
gladstone2026energybased,
title={Energy-Based Transformers are Scalable Learners and Thinkers},
author={Alexi Gladstone and Ganesh Nanduru and Md Mofijul Islam and Peixuan Han and Hyeonjeong Ha and Aman Chadha and Yilun Du and Heng Ji and Jundong Li and Tariq Iqbal},
booktitle={The Fourteenth International Conference on Learning Representations},
year={2026},
url={https://openreview.net/forum?id=ZBj3Qp1bYg}
}

@inproceedings{
janson2026stabilizing,
title={Stabilizing Native Low-Rank {LLM} Pretraining},
author={Paul Janson and Edouard Oyallon and Eugene Belilovsky},
booktitle={Forty-third International Conference on Machine Learning},
year={2026},
url={https://openreview.net/forum?id=kb5wjkqu8q}
}

@inproceedings{zellers2019hellaswag,
    title = "{H}ella{S}wag: Can a Machine Really Finish Your Sentence?",
    author = "Zellers, Rowan  and
      Holtzman, Ari  and
      Bisk, Yonatan  and
      Farhadi, Ali  and
      Choi, Yejin",
    editor = "Korhonen, Anna  and
      Traum, David  and
      M{\`a}rquez, Llu{\'i}s",
    booktitle = "Proceedings of the 57th Annual Meeting of the {A}ssociation for {C}omputational {L}inguistics",
    month = jul,
    year = "2019",
    address = "Florence, Italy",
    publisher = "Association for Computational Linguistics",
    url = "https://aclanthology.org/P19-1472/",
    doi = "10.18653/v1/P19-1472",
    pages = "4791--4800",
}

@inproceedings{bisk2019piqa,
  title={PIQA: Reasoning about Physical Commonsense in Natural Language},
  author={Yonatan Bisk and Rowan Zellers and Ronan Le Bras and Jian-Feng Gao and Yejin Choi},
  booktitle={AAAI Conference on Artificial Intelligence},
  year={2019},
  url={https://api.semanticscholar.org/CorpusID:208290939}
}
\bibliographystyle{mlsys2025}

\section*{AI Usage Disclosure}
Claude and ChatGPT assisted with manuscript editing to improve clarity and grammar. ChatGPT also assisted with the implementation of software functions and reviewed code for potential errors. These tools were not used to formulate the research questions, develop the proposed method, design the experiments, or interpret the empirical findings. The authors reviewed and verified all AI-assisted text and code and take responsibility for the final content.

\appendix

\clearpage
\begin{table*}[!t]
    \centering
    \caption{\textbf{134M model validation loss.} Final validation loss after training on 2.70B tokens, varying the number of replicas and selected peers at fixed density \(\delta=3.125\%\), with \topk = 128 and block size \(b=4096\). For each replica count, bold marks the best \textbf{decentralized} result and underlining marks the best \underline{sparse} result. For \ours, the diagonals trace paths of low validation loss degradation as the replica count increases.}
    \label{app:tab:134m}
    {\small
    \begin{tabular}{lccccccccc}
         \toprule
         \textbf{Method} & \textbf{Peers} $q$ & \textbf{Comm. Volume} & \textbf{P2P} & \textbf{Sparse} & \(\mathbf{R=8}\) & \(\mathbf{R=16}\) & \(\mathbf{R=32}\) & \(\mathbf{R=64}\) & \(\mathbf{R=128}\) \\
         \midrule
         AdamW DDP (\textit{ref.}) & gc & \(\nicefrac{2(R-1)}{R}~d\approx 2d\) & \xmark & \xmark & \multicolumn{5}{c}{3.17} \\
         \cdashline{1-10}
         \dadam & \(R-1\) & \((R-1)~d\) & \cmark & \xmark & 3.23 & 3.25 & 3.30 & 3.36 & 3.42 \\
         \gasloc & 1 & \(d\) & \cmark & \xmark & 3.20 & 3.23 & 3.28 & 3.35 & 3.44 \\
         \gasloc & 2 & \(2d\) & \cmark & \xmark & 3.19 & 3.21 & 3.24 & 3.31 & 3.37 \\
         \sparseloco & gc & \((R-1)\delta~d\) & \xmark & \cmark & 3.27 & 3.28 & 3.32 & 3.38 & 3.42 \\
         \choco & \(R-1\) & \((R-1)\delta~d\) & \cmark & \cmark & 3.27 & 3.35 & 3.47 & 3.67 & 3.86 \\
         \cdashline{1-10}
         \multirow{6}{*}{\oursbold} & 2 & \(2\delta~d=0.06d\) & \cmark & \cmark & 3.23 & 3.30 & 3.40 & 3.54 & 3.73 \\
         & 4 & \(4\delta~d=0.12d\) & \cmark & \cmark & \underline{\textbf{3.19}} & 3.23 & 3.31 & 3.39 & 3.52 \\
         & 8 & \(8\delta~d=0.25d\) & \cmark & \cmark & --- & \underline{\textbf{3.19}} & 3.24 & 3.30 & 3.39 \\
         & 16 & \(16\delta~d=0.5d\) & \cmark & \cmark & --- & --- & \underline{\textbf{3.20}} & 3.23 & 3.31 \\
         & 32 & \(32\delta~d=d\) & \cmark & \cmark & --- & --- & --- & \underline{\textbf{3.21}} & 3.25 \\
         & 64 & \(64\delta~d=2d\) & \cmark & \cmark & --- & --- & --- & --- & \underline{\textbf{3.23}} \\
         \bottomrule
    \end{tabular}
    }
\end{table*}

\section{Additional Results}

Tables~\ref{app:tab:134m} and~\ref{app:tab:551m} extend Table~\ref{tab:main-results} with validation losses for additional peer counts, providing the numerical results for the analysis of scaling across replicas.
Table~\ref{tab:training-times} supplements Figure~\ref{fig:training-speed} with absolute training times and additional combinations of peer count and message density.

\subsection{Expanding neighborhoods reduces degradation with replica count}
\label{app:sec:replicas}

We examine how the number of contacted peers affects training quality as the replica count increases. Tables~\ref{app:tab:134m} and~\ref{app:tab:551m} report validation losses across replica and peer counts at a fixed message density of 3.125\%. For the 134M model, increasing the replica count from 8 to 128 while keeping four peers raises validation loss by 10.3\%. Expanding the neighborhood according to \(q=R/2\) instead limits this increase to 1.3\%. At fixed message density, expanding the neighborhood therefore reduces the degradation observed when training is distributed across more replicas.

\begin{figure}[t]
    \centering
    \includegraphics[width=\linewidth]{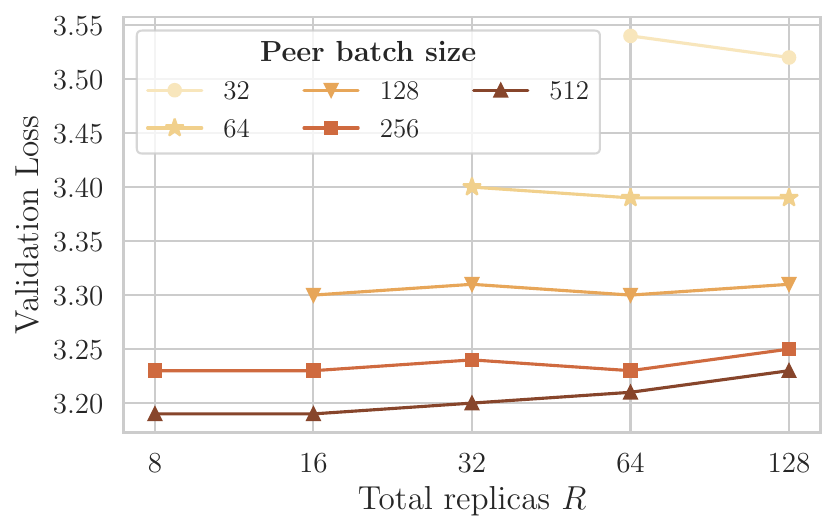}
    \caption{\textbf{Validation loss across replica counts at fixed peer batch size.} The 134M model is trained on 2.70B tokens at density \(\delta=3.125\%\). Each curve keeps the peer batch size \(qB/R\) constant, where \(q\) is the number of contacted peers and \(B=1024\) sequences is the global batch. Peer batch size excludes the replica's own batch. Expanding the neighborhood proportionally to replica count limits validation-loss degradation.}
    \label{app:fig:replicas}
\end{figure}

Figure~\ref{app:fig:replicas} groups configurations by their peer batch size, \(qB/R\), where \(B=1024\) sequences is the fixed global batch. Along each curve, the peer count grows proportionally to the replica count, keeping the total batch processed by the contacted peers constant while each replica's local batch shrinks. Validation loss remains approximately stable across these configurations, with larger peer batches consistently yielding lower loss. These results are consistent with expanding neighborhoods compensating for the shrinking local batch, although they do not isolate the mechanism responsible. In particular, peer batch size is not an effective optimization batch: peers update different local models, and only those transmitting a given coordinate contribute to its aggregation.

At peer batch sizes of 32 and 64 sequences, validation loss decreases slightly as the neighborhood expands. This suggests that peer batch size alone does not fully determine quality. The trend is consistent with Figure~\ref{fig:density-volume}, where allocating a fixed communication volume to more peers with sparser messages yields lower validation loss.

The 551M results in Table~\ref{app:tab:551m} confirm at a larger model scale that expanding the neighborhood reduces validation-loss degradation as the replica count increases.

\begin{table*}[t]
    \centering
    \caption{\textbf{551M model validation loss.} Final validation loss after training on 11.01B tokens, varying the number of replicas and selected peers at fixed density \(\delta=3.125\%\), with \topk = 128 and block size \(b=4096\). For each replica count, \textbf{bold} marks the best decentralized result and \underline{underlining} marks the best sparse result. These results confirm  at a larger model scale the diagonal trend observed for \ours at 134M.}
    \label{app:tab:551m}
    {\small
    \begin{tabular}{lccccccc}
         \toprule
         \textbf{Method} & \textbf{Peers} $q$ & \textbf{Comm. Volume} & \textbf{P2P} & \textbf{Sparse} & \(\mathbf{R=8}\) & \(\mathbf{R=16}\) & \(\mathbf{R=32}\) \\
         \midrule
         AdamW DDP (\textit{ref.}) & gc & \(\nicefrac{2(R-1)}{R}~d\approx 2d\) & \xmark & \xmark & \multicolumn{3}{c}{2.63} \\
         \cdashline{1-8}
         \dadam & \(R-1\) & \((R-1)~d\) & \cmark & \xmark & 2.70 & 2.73 & 2.77 \\
         \gasloc & 1 & \(d\) & \cmark & \xmark & 2.67 & 2.71 & 2.75 \\
         \gasloc & 2 & \(2d\) & \cmark & \xmark & \textbf{2.64} & 2.68 & 2.73 \\
         \sparseloco & gc & \((R-1) \delta~d\) & \xmark & \cmark & 2.71 & 2.74 & 2.78 \\
         \choco & \(R-1\) & \((R-1)\delta~d\) & \cmark & \cmark & 2.73 & 2.81 & 2.95 \\
         \cdashline{1-8}
         \multirow{4}{*}{\oursbold} & 2 & \(2\delta~d=0.06d\) & \cmark & \cmark & 2.71 & 2.75 & 2.77 \\
         & 4 & \(4\delta~d=0.12d\) & \cmark & \cmark & \underline{2.65} & 2.71 & 2.74 \\
         & 8 & \(8\delta~d=0.25d\) & \cmark & \cmark & --- & \underline{\textbf{2.68}} & 2.71 \\
         & 16 & \(16\delta~d=0.5d\) & \cmark & \cmark & --- & --- & \underline{\textbf{2.70}} \\
         \bottomrule
    \end{tabular}
    }
\end{table*}

\begin{table*}[t]
    \centering
    \caption{\textbf{End-to-end training times for the 134M and 551M models.} Times are reported as hours:minutes and rounded to the nearest minute. Each model group shows the uncapped NCCL setting and bandwidth caps. The lowest time per model size and bandwidth is highlighted in \textbf{bold}.}
    \label{tab:training-times}
    \resizebox{\linewidth}{!}{
    \begin{tabular}{lccccccccccccc}
        \toprule
        &&& \multicolumn{5}{c}{\textbf{134M}} & \multicolumn{5}{c}{\textbf{551M}} \\
        \cmidrule(lr){4-8} \cmidrule(lr){9-13}
        \textbf{Method} & \textbf{Peers} $q$ & \textbf{Density $\delta$} & \shortstack{\textbf{NCCL} \\ \textbf{(no cap)}} & \shortstack{\textbf{10} \\ \textbf{Gbps}} & \shortstack{\textbf{5} \\ \textbf{Gbps}} & \shortstack{\textbf{2} \\ \textbf{Gbps}} & \shortstack{\textbf{1} \\ \textbf{Gbps}} & \shortstack{\textbf{NCCL} \\ \textbf{(no cap)}} & \shortstack{\textbf{10} \\ \textbf{Gbps}} & \shortstack{\textbf{5} \\ \textbf{Gbps}} & \shortstack{\textbf{2} \\ \textbf{Gbps}} & \shortstack{\textbf{1} \\ \textbf{Gbps}} \\
        \midrule
        DDP & --- & --- & 2:05 & 4:49 & 5:14 & 5:35 & 7:08 & 23:24 & 48:40 & 51:56 & 55:16 & 69:38 \\
        \hline
        \gasloc & 1 & --- & 1:38 & 2:24 & 2:27 & 2:32 & 4:03 & 21:49 & 34:28 & 34:54 & 35:54 & 57:04 \\
        \gasloc & 2 & --- & 1:47 & 3:57 & 4:18 & 4:33 & 6:01 & 24:51 & 58:22 & 63:13 & 67:26 & 94:20 \\
        \hline
        \multirow{4}{*}{\oursbold} & 2 & 3.125\% & \textbf{1:34} & \textbf{1:38} & \textbf{1:40} & \textbf{1:40} & \textbf{1:43} & \textbf{19:56} & \textbf{21:11} & \textbf{21:21} & \textbf{21:30} & \textbf{22:19} \\
        & 4 & 3.125\% & 1:35 & 1:46 & 1:48 & 1:50 & 1:55 & 20:12 & 23:17 & 23:39 & 24:17 & 25:56 \\
        & 2 & 6.25\% & 1:34 & 1:44 & 1:46 & 1:46 & 1:53 & 20:15 & 22:48 & 23:12 & 23:26 & 25:09 \\
        & 4 & 6.25\% & 1:36 & 1:59 & 2:04 & 2:07 & 2:17 & 20:25 & 26:46 & 27:54 & 29:13 & 32:20 \\
        \bottomrule
    \end{tabular}
    }
\end{table*}

\section{Reproducibility}

\subsection{Peer sampling}
\label{app:peer-sampling}
The \textsc{sample\_peers} procedure in Algorithm~\ref{alg:sparse-gasloc} is implemented as follows.
At each round $t$, every replica independently generates the same uniform permutation $\sigma^{(t)}$ of the $R$ replicas using a shared seed and the round index $t$.
This permutation arranges replicas on a cycle without communication to coordinate peer selection.
For an even peer count $q<R$, each replica exchanges messages with the $q/2$ replicas preceding it and the $q/2$ replicas following it, with positions taken modulo $R$.
For the one-peer GASLoC configuration, R must be even, and consecutive replicas in the permutation form disjoint pairs: the first exchanges with the second, the third with the fourth, and so on.
Both constructions give each replica exactly $q$ distinct peers and make every connection symmetric.
Generating a fresh permutation at each round changes the communication partners while preserving the peer count, allowing information to propagate beyond a fixed neighborhood.

\subsection{Software and hardware setup}
The training simulator uses Python 3.12, PyTorch 2.10.0, and TorchTitan 0.2.1.
The implementation used for timing measurements is based on TorchTitan 0.2.2 and the June 23, 2026 nightly release of \texttt{torchft}.
All models use the LLaMA-2 tokenizer with a vocabulary of 32K tokens, available from the \href{https://huggingface.co/togethercomputer/LLaMA-2-7B-32K/blob/main/tokenizer.model}{Together AI repository}.

Training runs for the 134M and 551M models use compute nodes with four NVIDIA H100 GPUs per node, connected through NVLink within each node and InfiniBand across nodes.
These runs simulate multiple logical replicas on each GPU.
OLMoE training runs use a single node with eight NVIDIA H200 GPUs connected through NVLink.
Training times and throughput are measured separately on a single node with eight NVIDIA RTX A6000 GPUs, with one physical training replica per GPU.

\subsection{Communication bandwidth limits}
\label{app:sec:bandwidth-limits}

For the global bandwidth experiments, each of the eight GPU replicas runs in a separate Linux network namespace connected through a virtual Ethernet pair to a private bridge.
Linux traffic control \texttt{tc} applies the same fixed bandwidth cap to every replica, independently limiting its aggregate outgoing and incoming traffic.
The outgoing cap is installed on the namespace interface, and the incoming cap on the host end of its virtual Ethernet pair, using \texttt{netem} with a queue limit of 10,000 packets and packet offloads disabled.
All peer connections of a replica share these caps. NCCL uses IPv4 sockets on the virtual interface, with any other GPU transport technology disabled to prevent communication from bypassing the limits.
Control and monitoring traffic also share the capped interfaces.

For the selective bandwidth experiment, we compare DDP and \ours using the same namespace isolation and assign replicas 1--3 to island A, 4--6 to island B, and 7--8 to island C (\cref{fig:islands}).
After five minutes of unrestricted communication, a host controller caps only traffic between islands A and B at 500~Mbit/s (or 100~Mbit/s) for five minutes, then removes the caps for five minutes of recovery.
Each source interface applies a separate bandwidth cap to each affected peer based on its destination IP address.
The cap applies independently in each direction, while all other worker pairs remain unrestricted.
Transitions follow a common monotonic host clock independently of training progress.
DDP uses TorchFT with NCCL sockets, while \ours exchanges sparse payloads with four rotating peers through a TCP transport that stages tensors in CPU memory.
A calibration during the initial healthy phase sets the exchange deadline of the windowed version of \ours to twice the pooled 99th percentile of exchange duration.
Only complete payloads with matching attempt and schema identities received before the deadline contribute to the update, allowing communication with healthy peers to remain useful when capped transfers miss the deadline.

\subsection{Random seeds and reporting precision}

We evaluate each training configuration using a single random seed.
Recent pretraining studies likewise report single-seed results at selected model scales, citing the computational cost of repeated runs~\cite{loeschcke2024loqt,gladstone2026energybased}.
Given this cost, we prioritize allocating our computational budget to hyperparameter tuning rather than repeated training across seeds.
Related distributed pretraining studies also report their main comparisons without across-seed uncertainty estimates~\cite{douillard2023diloco,kolehmainen2026noloco,janson2026stabilizing}.

As a closely related reference, \gasloc evaluates 134M-parameter models with 16 replicas over four random seeds, reporting validation-loss standard deviations of at most 0.011 across the evaluated methods and configurations~\cite{cagnasso2026unifying}.
Following its main-result reporting convention, we report validation losses rounded to two decimal places, evaluated on a held-out set of 4M tokens for the 134M and 551M models.

\subsection{Hyperparameters}

We select hyperparameters by grid search using validation loss on a second separate evaluation set of 1M tokens.
For the 134M model, we search the peak inner learning rate over $\{1.4,2,2.8,4,5,6\}\times10^{-3}$, the outer learning rate over $\{0.8,0.9,1.0\}$, and the outer Nesterov momentum over $\{0.1,0.2,0.3\}$.
For the 551M model, the corresponding learning rate grids are $\{3,4,5,6\}\times10^{-3}$ and $\{0.8,0.9\}$, with the same outer momentum grid.
The inner learning rate warms up from one thousandth of its peak over the first 10\% of training steps, followed by cosine decay to 10\% of the peak.
The inner optimizer is AdamW with $\beta_1=0.9$.
Following \citet{belilovsky2026how}, we set $\beta_2$ to $0.95$, $0.975$, $0.987$, $0.994$, and $0.997$ for $R=8,16,32,64,128$, respectively.
For OLMoE, after short preliminary runs to assess initial training stability, we fix the peak inner learning rate to $2\times10^{-3}$, with warmup over the first 5\% of training steps and the same initial and final learning rate fractions.
We use $\beta_1=\beta_2=0.9$, an outer learning rate of $0.9$, and outer Nesterov momentum of $0.2$.

Across all models, AdamW uses $\epsilon=10^{-8}$ and weight decay of $0.1$.
Inner gradients are clipped to a norm of $1.0$, while the outer optimizer uses no gradient clipping and a constant learning rate.
Unless specified otherwise, the \topk block size is fixed to $b=4096$, and the error feedback decay factor is fixed to $0.95$ following \citet{sarfi2026sparseloco}.
\Cref{tab:hps} reports the selected hyperparameters.

\begin{table*}[!t]
    \centering
    \caption{Selected hyperparameters for \ours on the 134M and 551M models across replica counts.}
    \label{tab:hps}
    \begin{tabular}{lccccccc}
        \toprule
        \textbf{Model} & \textbf{Peers} $q$ & \textbf{HP} & $\mathbf{R=8}$ & $\mathbf{R=16}$ & $\mathbf{R=32}$ & $\mathbf{R=64}$ & $\mathbf{R=128}$ \\
        \midrule
        \multirow{18}{*}{134M} & \multirow{3}{*}{2} & $\eta_{in}$ & 2e-3 & 2e-3 & 2e-3 & 2e-3 & 2e-3 \\
        && $\eta_{out}$ & 0.9 & 0.9 & 0.9 & 0.8 & 0.9 \\
        && $\mu_{out}$ & 0.2 & 0.3 & 0.3 & 0.3 & 0.3 \\
        \cline{2-8}
        & \multirow{3}{*}{4} & $\eta_{in}$ & 4e-3 & 2.8e-3 & 2e-3 & 2e-3 & 2e-3 \\
        && $\eta_{out}$ & 0.9 & 0.9 & 0.9 & 0.9 & 0.9 \\
        && $\mu_{out}$ & 0.2 & 0.3 & 0.3 & 0.3 & 0.3 \\
        \cline{2-8}
        & \multirow{3}{*}{8} & $\eta_{in}$ & \multirow{3}{*}{---} & 2.8e-3 & 2e-3 & 2e-3 & 2e-3 \\
        && $\eta_{out}$ &  & 0.9 & 0.8 & 0.9 & 0.9 \\
        && $\mu_{out}$ &  & 0.3 & 0.3 & 0.3 & 0.3 \\
        \cline{2-8}
        & \multirow{3}{*}{16} & $\eta_{in}$ & \multirow{3}{*}{---} & \multirow{3}{*}{---} & 2e-3 & 2.8e-3 & 2.8e-3 \\
        && $\eta_{out}$ &  &  & 0.9 & 0.9 & 0.9 \\
        && $\mu_{out}$ &  &  & 0.3 & 0.3 & 0.2 \\
        \cline{2-8}
        & \multirow{3}{*}{32} & $\eta_{in}$ & \multirow{3}{*}{---} & \multirow{3}{*}{---} & \multirow{3}{*}{---} & 4e-3 & 4e-3 \\
        && $\eta_{out}$ &  &  &  & 0.9 & 0.8 \\
        && $\mu_{out}$ &  &  &  & 0.2 & 0.3 \\
        \cline{2-8}
        & \multirow{3}{*}{64} & $\eta_{in}$ & \multirow{3}{*}{---} & \multirow{3}{*}{---} & \multirow{3}{*}{---} & \multirow{3}{*}{---} & 5e-3 \\
        && $\eta_{out}$ &  &  &  & & 0.9 \\
        && $\mu_{out}$ &  &  &  & & 0.2 \\
        \midrule
        \multirow{12}{*}{551M} & \multirow{3}{*}{2} & $\eta_{in}$ & 4e-3 & 4e-3 & 3e-3 & \multirow{3}{*}{---} & \multirow{3}{*}{---} \\
        && $\eta_{out}$ & 0.9 & 0.9 & 0.9 & & \\
        && $\mu_{out}$ & 0.3 & 0.3 & 0.3 & & \\
        \cline{2-8}
        & \multirow{3}{*}{4} & $\eta_{in}$ & 5e-3 & 5e-3 & 4e-3 & \multirow{3}{*}{---} & \multirow{3}{*}{---} \\
        && $\eta_{out}$ & 0.9 & 0.9 & 0.9 & & \\
        && $\mu_{out}$ & 0.1 & 0.2 & 0.2 &  & \\
        \cline{2-8}
        & \multirow{3}{*}{8} & $\eta_{in}$ & \multirow{3}{*}{---} & 6e-3 & 5e-3 & \multirow{3}{*}{---} & \multirow{3}{*}{---} \\
        && $\eta_{out}$ &  & 0.9 & 0.9 & & \\
        && $\mu_{out}$ &  & 0.2 & 0.2 &  &  \\
        \cline{2-8}
        & \multirow{3}{*}{16} & $\eta_{in}$ & \multirow{3}{*}{---} & \multirow{3}{*}{---} & 5e-3 & \multirow{3}{*}{---} & \multirow{3}{*}{---} \\
        && $\eta_{out}$ &  &  & 0.9 & & \\
        && $\mu_{out}$ &  &  & 0.1 &  &  \\
        \bottomrule
    \end{tabular}
\end{table*}

\end{document}